%% file: ijphp.tex
\documentclass[zpreprint,zbstdefault,british]{./LaTeX/zeus/zeus_K2.9paper}
\usepackage{./LaTeX/zeus/zeusdet_units}
\usepackage{./LaTeX/styles/lineno}
\input ./LaTeX/zeus/zeus_defs.tex
\input ijphp-cit.tex
\begin{document}
\input{ijphp-tit}
\thispagestyle{empty}
\cleardoublepage
\input{ijphp-aut}
\cleardoublepage
\pagenumbering{arabic}
%
%
\pagestyle{scrheadings}
\include{ijphp-txt}
\include{ijphp-ref}
\include{ijphp-tab}
\include{ijphp-fig}
%
%
\end{document}

%% file: LaTeX/zeus/zeus_defs.tex
\chardef\usc=95
\chardef\til=126
\catcode`\@=11 
\DeclareRobustCommand\xdotspace{\futurelet\@let@token\@xdotspace}
\def\@xdotspace{%
  \ifx\@let@token.\else
  \ifx\@let@token\bgroup.\else
  \ifx\@let@token\egroup.\else
  \ifx\@let@token\/.\else
  \ifx\@let@token\ .\else
  \ifx\@let@token~.\else
  \ifx\@let@token!.\else
  \ifx\@let@token,.\else
  \ifx\@let@token:.\else
  \ifx\@let@token;.\else
  \ifx\@let@token?.\else
  \ifx\@let@token/.\else
  \ifx\@let@token'.\else
  \ifx\@let@token).\else
  \ifx\@let@token-.\else
  \ifx\@let@token\@xobeysp.\else
  \ifx\@let@token\space.\else
  \ifx\@let@token\@sptoken.\else
   .\space
   \fi\fi\fi\fi\fi\fi\fi\fi\fi\fi\fi\fi\fi\fi\fi\fi\fi\fi}
\catcode`\@=12 

\newcommand{\stru}[2]{%
   \relax\ifmmode\hbox{\vrule height#1 depth#2 width0pt}%
   \else\vrule height#1 depth#2 width0pt\fi}

\newcommand{\Ronum}[1]{\uppercase\expandafter{\romannumeral#1}}
\newcommand{\ronum}[1]{\expandafter{\romannumeral#1}}
\DeclareRobustCommand{\LaTeXZ}{%
  \LaTeX\kern-.05em4\kern-.1em
  {\raisebox{-0.2ex}{$\scriptstyle\text{ZEUS}$}}\xspace}

\newcommand{\slashfrac}[2]{%
  \raisebox{0.5ex}{\ensuremath #1}\kern-0.12em/\kern-0.08em
  \raisebox{-.8ex}{\ensuremath #2}}

\newcommand{\sqr}[3]{%
    {\vcenter{\hrule height.#3ex\hbox{\vrule width.#2ex height#1ex
     \kern#1ex\vrule width.#3ex}\hrule height.#2ex}}}

\catcode`\@=11 
\newcommand{\parenbar}{\mathpalette\p@renb@r}
\def\p@renb@r#1#2{\vbox{%
  \ifx#1\scriptscriptstyle \dimen@.7em\dimen@ii.2em\else
  \ifx#1\scriptstyle \dimen@.8em\dimen@ii.25em\else
  \dimen@1em\dimen@ii.4em\fi\fi \offinterlineskip
  \ialign{\hfill##\hfill\cr
    \vbox{\hrule width\dimen@ii}\cr
    \noalign{\vskip-.3ex}%
    \hbox to\dimen@{$\mathchar300\hfil\mathchar301$}\cr
    \noalign{\vskip-.3ex}%
    $#1#2$\cr}}}
\catcode`\@=12 

\ifzeusunit
  
\else
  
\fi

\newcommand{\IP}{{\rm I$\kern-0.01667em$P}\xspace}

\mathchardef\qsm=63
\mathchardef\pls=43
\mathchardef\mns=512
\mathchardef\plm=518
\mathchardef\eql=61
\mathchardef\smallleft=300
\mathchardef\smallright=301
\mathchardef\les=316
\mathchardef\gre=318
\mathchardef\leq=532
\mathchardef\grq=533

\catcode`\@=11 
\newcounter{pict@width}
\newcounter{pict@height}
\newlength{\pict@scale}
\newcommand{\psfigadd}[4]{%
\setcounter{pict@width}{1*\ratio{#2+\pict@scale/2}{\pict@scale}}
\setcounter{pict@height}{1*\ratio{#3+\pict@scale/2}{\pict@scale}}
\setlength{\unitlength}{\pict@scale}
\hbox to #2{\hspace{-\fill}\begin{picture}(\thepict@width,\thepict@height)
\put(0,0){\psfig{figure=#1,width=#2,height=#3,clip=}}
\SetScale{0.283466457}
\SetWidth{1.763889}
{#4}
\end{picture}}
}
\newcounter{pict@widthfst}
\newcounter{pict@widthscd}
\newcounter{pict@widthtot}
\newcommand{\psfigaddtwo}[7]{%
\setcounter{pict@widthfst}{1*\ratio{#2+\pict@scale/2}{\pict@scale}}
\setcounter{pict@widthscd}{1*\ratio{#2+#4+\pict@scale/2}{\pict@scale}}
\setcounter{pict@widthtot}{1*\ratio{#2+#4+#6+\pict@scale/2}{\pict@scale}}
\setcounter{pict@height}{1*\ratio{#3+\pict@scale/2}{\pict@scale}}
\setlength{\unitlength}{\pict@scale}
\hbox{\hspace{-\fill}\begin{picture}(\thepict@widthtot,\thepict@height)
\put(0,0){\psfig{figure=#1,width=#2,height=#3,clip=}}
\put(\thepict@widthscd,0){\psfig{figure=#5,width=#6,height=#3,clip=}}
\SetScale{0.283466457}
\SetWidth{1.763889}
{#7}
\end{picture}}
}
\newcommand{\psfigror}[4]{%
\setcounter{pict@width}{1*\ratio{#2+\pict@scale/2}{\pict@scale}}
\setcounter{pict@height}{1*\ratio{#3+\pict@scale/2}{\pict@scale}}
\setlength{\unitlength}{\pict@scale}
\hbox{\begin{picture}(\thepict@width,\thepict@height)
\put(0,\thepict@height){\psfig{figure=#1,width=#3,height=#2,clip=,angle=270}}
\SetScale{0.283466457}
\SetWidth{1.763889}
{#4}
\end{picture}}
}
\newcommand{\psfigrol}[4]{%
\setcounter{pict@width}{1*\ratio{#2+\pict@scale/2}{\pict@scale}}
\setcounter{pict@height}{1*\ratio{#3+\pict@scale/2}{\pict@scale}}
\setlength{\unitlength}{\pict@scale}
\hbox{\begin{picture}(\thepict@width,\thepict@height)
\put(0,0){\psfig{figure=#1,width=#3,height=#2,clip=,angle=90}}
\SetScale{0.283466457}
\SetWidth{1.763889}
{#4}
\end{picture}}
}
\catcode`\@=12 
\newlength\listtextwidth

\catcode`\@=11 
\newlength{\@tabfninsert}
\newlength{\@tabfnwidth}
\newcommand{\tabfootnote}[2]{%
  \setlength{\@tabfninsert}{0.8em}
  \setlength{\@tabfnwidth}{\textwidth}
  \addtolength{\@tabfnwidth}{-\@tabfninsert}
  \addtolength{\@tabfnwidth}{-0.4em}
  \noindent\makebox[\@tabfninsert][r]{\footnotesize$^{#1}$\hfil}\hfill%
  \parbox[t]{\@tabfnwidth}{\footnotesize #2\hfill}}
\catcode`\@=12 

%% file: ijphp-cit.tex
\def\citeCTD{{\cite{%
nim:a279:290,*npps:b32:181,*nim:a338:254%
}}\xspace}
\def\citeMVD{{\cite{%
nim:a581:656%
}}\xspace}
\def\citeSTT{{\cite{%
nim:a535:191%
}}\xspace}
\def\citeCAL{{\cite{%
nim:a309:77,*nim:a309:101,*nim:a321:356,*nim:a336:23%
}}\xspace}
\def\citeHES{{\cite{%
nim:a277:176%
}}\xspace}
\def\citeSixmT{{\cite{%
thesis:gosau:2007%
}}\xspace}
\def\citeBAC{{\cite{%
nim:a300:480%
}}\xspace}
\def\citePCAL{{\cite{%
desy-92-066,*zfp:c63:391,*acpp:b32:2025%
}}\xspace}
\def\citeSPECTRO{{\cite{%
nim:a565:572%
}}\xspace}

%% file: ijphp-tit.tex
%
%
\prepnum{}
\prepdate{}

\zeustitle{
Measurement of inelastic \(J/\psi\) and \(\psi^{\prime}\) 
photoproduction at HERA
}

\zeusauthor{ZEUS Collaboration}
\draftversion{}
\zeusdate{}

\maketitle

\begin{abstract} \noindent
The cross sections for inelastic photoproduction of \(J/\psi\) 
and \(\psi^{\prime}\) mesons have been measured in $ep$ collisions 
with the ZEUS detector at HERA, using an integrated luminosity of 
468 $ \pbi $ collected in the period 1996--2007.
The \(\psi^{\prime}\) to  \(J/\psi\) cross section 
ratio was measured in the range \(0.55 < z < 0.9\) and 
\(60 < W < 190 \gev\) as a function of $W$, $z$ and $p_T$.
Here \(W \) denotes the photon-proton centre-of-mass energy,
\(z\) is the fraction of the incident photon energy 
carried by the meson and $p_T$ is the transverse momentum of the 
meson with respect to the beam axis.
The \(J/\psi\) cross sections were measured for \(0.1 < z < 0.9\), 
\(60 < W < 240 \gev\) and \(p_T > 1 \gev\).
Theoretical predictions within the 
non-relativistic QCD framework including NLO colour--singlet and 
colour--octet contributions were compared to the data, as were
predictions based on the \(k_T\)--factorisation approach.
\end{abstract}

%% file: ijphp-aut.tex
%
%
%
%

                                                   %
\begin{center}
{                      \Large  The ZEUS Collaboration              }
\end{center}

{\small


        {\raggedright
H.~Abramowicz$^{45, ah}$, 
I.~Abt$^{35}$, 
L.~Adamczyk$^{13}$, 
M.~Adamus$^{54}$, 
R.~Aggarwal$^{7, c}$, 
S.~Antonelli$^{4}$, 
P.~Antonioli$^{3}$, 
A.~Antonov$^{33}$, 
M.~Arneodo$^{50}$, 
O.~Arslan$^{5}$
% ...
        }

\newpage


\makebox[3em]{$^{1}$}
\begin{minipage}[t]{14cm}
{\it Argonne National Laboratory, Argonne, Illinois 60439-4815, USA}~$^{A}$

\end{minipage}\\
\makebox[3em]{$^{2}$}
\begin{minipage}[t]{14cm}
{\it Andrews University, Berrien Springs, Michigan 49104-0380, USA}

\end{minipage}\\
\makebox[3em]{$^{3}$}
\begin{minipage}[t]{14cm}
{\it INFN Bologna, Bologna, Italy}~$^{B}$

\end{minipage}\\
\makebox[3em]{$^{4}$}
\begin{minipage}[t]{14cm}
{\it University and INFN Bologna, Bologna, Italy}~$^{B}$

\end{minipage}\\
\makebox[3em]{$^{5}$}
\begin{minipage}[t]{14cm}
{\it Physikalisches Institut der Universit\"at Bonn,
Bonn, Germany}~$^{C}$

\end{minipage}\\
\makebox[3em]{$^{6}$}
\begin{minipage}[t]{14cm}
{\it H.H.~Wills Physics Laboratory, University of Bristol,
Bristol, United Kingdom}~$^{D}$

\end{minipage}\\
\makebox[3em]{$^{7}$}
\begin{minipage}[t]{14cm}
{\it Panjab University, Department of Physics, Chandigarh, India}

\end{minipage}\\
\makebox[3em]{$^{8}$}
\begin{minipage}[t]{14cm}
{\it Calabria University,
Physics Department and INFN, Cosenza, Italy}~$^{B}$

\end{minipage}\\
\makebox[3em]{$^{9}$}
\begin{minipage}[t]{14cm}
{\it Institute for Universe and Elementary Particles, Chonnam National University,\\
Kwangju, South Korea}

\end{minipage}\\
\makebox[3em]{$^{10}$}
\begin{minipage}[t]{14cm}
{\it Jabatan Fizik, Universiti Malaya, 50603 Kuala Lumpur, Malaysia}~$^{E}$

\end{minipage}\\
\makebox[3em]{$^{11}$}
\begin{minipage}[t]{14cm}
{\it Nevis Laboratories, Columbia University, Irvington on Hudson,
New York 10027, USA}~$^{F}$

\end{minipage}\\
\makebox[3em]{$^{12}$}
\begin{minipage}[t]{14cm}
{\it The Henryk Niewodniczanski Institute of Nuclear Physics, Polish Academy of \\
Sciences, Krakow, Poland}~$^{G}$

\end{minipage}\\
\makebox[3em]{$^{13}$}
\begin{minipage}[t]{14cm}
{\it AGH-University of Science and Technology, Faculty of Physics and Applied Computer
Science, Krakow, Poland}~$^{H}$

\end{minipage}\\
\makebox[3em]{$^{14}$}
\begin{minipage}[t]{14cm}
{\it Department of Physics, Jagellonian University, Cracow, Poland}

\end{minipage}\\
\makebox[3em]{$^{15}$}
\begin{minipage}[t]{14cm}
{\it Deutsches Elektronen-Synchrotron DESY, Hamburg, Germany}

\end{minipage}\\
\makebox[3em]{$^{16}$}
\begin{minipage}[t]{14cm}
{\it Deutsches Elektronen-Synchrotron DESY, Zeuthen, Germany}

\end{minipage}\\
\makebox[3em]{$^{17}$}
\begin{minipage}[t]{14cm}
{\it INFN Florence, Florence, Italy}~$^{B}$

\end{minipage}\\
\makebox[3em]{$^{18}$}
\begin{minipage}[t]{14cm}
{\it University and INFN Florence, Florence, Italy}~$^{B}$

\end{minipage}\\
\makebox[3em]{$^{19}$}
\begin{minipage}[t]{14cm}
{\it Fakult\"at f\"ur Physik der Universit\"at Freiburg i.Br.,
Freiburg i.Br., Germany}

\end{minipage}\\
\makebox[3em]{$^{20}$}
\begin{minipage}[t]{14cm}
{\it School of Physics and Astronomy, University of Glasgow,
Glasgow, United Kingdom}~$^{D}$

\end{minipage}\\
\makebox[3em]{$^{21}$}
\begin{minipage}[t]{14cm}
{\it Department of Engineering in Management and Finance, Univ. of
the Aegean, Chios, Greece}

\end{minipage}\\
\makebox[3em]{$^{22}$}
\begin{minipage}[t]{14cm}
{\it Hamburg University, Institute of Experimental Physics, Hamburg,
Germany}~$^{I}$

\end{minipage}\\
\makebox[3em]{$^{23}$}
\begin{minipage}[t]{14cm}
{\it Imperial College London, High Energy Nuclear Physics Group,
London, United Kingdom}~$^{D}$

\end{minipage}\\
\makebox[3em]{$^{24}$}
\begin{minipage}[t]{14cm}
{\it Institute of Particle and Nuclear Studies, KEK,
Tsukuba, Japan}~$^{J}$

\end{minipage}\\
\makebox[3em]{$^{25}$}
\begin{minipage}[t]{14cm}
{\it Institute of Physics and Technology of Ministry of Education and
Science of Kazakhstan, Almaty, Kazakhstan}

\end{minipage}\\
\makebox[3em]{$^{26}$}
\begin{minipage}[t]{14cm}
{\it Institute for Nuclear Research, National Academy of Sciences, Kyiv, Ukraine}

\end{minipage}\\
\makebox[3em]{$^{27}$}
\begin{minipage}[t]{14cm}
{\it Department of Nuclear Physics, National Taras Shevchenko University of Kyiv, Kyiv, Ukraine}

\end{minipage}\\
\makebox[3em]{$^{28}$}
\begin{minipage}[t]{14cm}
{\it Kyungpook National University, Center for High Energy Physics, Daegu,
South Korea}~$^{K}$

\end{minipage}\\
\makebox[3em]{$^{29}$}
\begin{minipage}[t]{14cm}
{\it Institut de Physique Nucl\'{e}aire, Universit\'{e} Catholique de Louvain, Louvain-la-Neuve,\\
Belgium}~$^{L}$

\end{minipage}\\
\makebox[3em]{$^{30}$}
\begin{minipage}[t]{14cm}
{\it Departamento de F\'{\i}sica Te\'orica, Universidad Aut\'onoma
de Madrid, Madrid, Spain}~$^{M}$

\end{minipage}\\
\makebox[3em]{$^{31}$}
\begin{minipage}[t]{14cm}
{\it Department of Physics, McGill University,
Montr\'eal, Qu\'ebec, Canada H3A 2T8}~$^{N}$

\end{minipage}\\
\makebox[3em]{$^{32}$}
\begin{minipage}[t]{14cm}
{\it Meiji Gakuin University, Faculty of General Education,
Yokohama, Japan}~$^{J}$

\end{minipage}\\
\makebox[3em]{$^{33}$}
\begin{minipage}[t]{14cm}
{\it Moscow Engineering Physics Institute, Moscow, Russia}~$^{O}$

\end{minipage}\\
\makebox[3em]{$^{34}$}
\begin{minipage}[t]{14cm}
{\it Lomonosov Moscow State University, Skobeltsyn Institute of Nuclear Physics,
Moscow, Russia}~$^{P}$

\end{minipage}\\
\makebox[3em]{$^{35}$}
\begin{minipage}[t]{14cm}
{\it Max-Planck-Institut f\"ur Physik, M\"unchen, Germany}

\end{minipage}\\
\makebox[3em]{$^{36}$}
\begin{minipage}[t]{14cm}
{\it NIKHEF and University of Amsterdam, Amsterdam, Netherlands}~$^{Q}$

\end{minipage}\\
\makebox[3em]{$^{37}$}
\begin{minipage}[t]{14cm}
{\it Physics Department, Ohio State University,
Columbus, Ohio 43210, USA}~$^{A}$

\end{minipage}\\
\makebox[3em]{$^{38}$}
\begin{minipage}[t]{14cm}
{\it Department of Physics, University of Oxford,
Oxford, United Kingdom}~$^{D}$

\end{minipage}\\
\makebox[3em]{$^{39}$}
\begin{minipage}[t]{14cm}
{\it INFN Padova, Padova, Italy}~$^{B}$

\end{minipage}\\
\makebox[3em]{$^{40}$}
\begin{minipage}[t]{14cm}
{\it Dipartimento di Fisica dell' Universit\`a and INFN,
Padova, Italy}~$^{B}$

\end{minipage}\\
\makebox[3em]{$^{41}$}
\begin{minipage}[t]{14cm}
{\it Department of Physics, Pennsylvania State University, University Park,\\
Pennsylvania 16802, USA}~$^{F}$

\end{minipage}\\
\makebox[3em]{$^{42}$}
\begin{minipage}[t]{14cm}
{\it Polytechnic University, Tokyo, Japan}~$^{J}$

\end{minipage}\\
\makebox[3em]{$^{43}$}
\begin{minipage}[t]{14cm}
{\it Dipartimento di Fisica, Universit\`a 'La Sapienza' and INFN,
Rome, Italy}~$^{B}$

\end{minipage}\\
\makebox[3em]{$^{44}$}
\begin{minipage}[t]{14cm}
{\it Rutherford Appleton Laboratory, Chilton, Didcot, Oxon,
United Kingdom}~$^{D}$

\end{minipage}\\
\makebox[3em]{$^{45}$}
\begin{minipage}[t]{14cm}
{\it Raymond and Beverly Sackler Faculty of Exact Sciences, School of Physics, \\
Tel Aviv University, Tel Aviv, Israel}~$^{R}$

\end{minipage}\\
\makebox[3em]{$^{46}$}
\begin{minipage}[t]{14cm}
{\it Department of Physics, Tokyo Institute of Technology,
Tokyo, Japan}~$^{J}$

\end{minipage}\\
\makebox[3em]{$^{47}$}
\begin{minipage}[t]{14cm}
{\it Department of Physics, University of Tokyo,
Tokyo, Japan}~$^{J}$

\end{minipage}\\
\makebox[3em]{$^{48}$}
\begin{minipage}[t]{14cm}
{\it Tokyo Metropolitan University, Department of Physics,
Tokyo, Japan}~$^{J}$

\end{minipage}\\
\makebox[3em]{$^{49}$}
\begin{minipage}[t]{14cm}
{\it Universit\`a di Torino and INFN, Torino, Italy}~$^{B}$

\end{minipage}\\
\makebox[3em]{$^{50}$}
\begin{minipage}[t]{14cm}
{\it Universit\`a del Piemonte Orientale, Novara, and INFN, Torino,
Italy}~$^{B}$

\end{minipage}\\
\makebox[3em]{$^{51}$}
\begin{minipage}[t]{14cm}
{\it Department of Physics, University of Toronto, Toronto, Ontario,
Canada M5S 1A7}~$^{N}$

\end{minipage}\\
\makebox[3em]{$^{52}$}
\begin{minipage}[t]{14cm}
{\it Physics and Astronomy Department, University College London,
London, United Kingdom}~$^{D}$

\end{minipage}\\
\makebox[3em]{$^{53}$}
\begin{minipage}[t]{14cm}
{\it Faculty of Physics, University of Warsaw, Warsaw, Poland}

\end{minipage}\\
\makebox[3em]{$^{54}$}
\begin{minipage}[t]{14cm}
{\it National Centre for Nuclear Research, Warsaw, Poland}

\end{minipage}\\
\makebox[3em]{$^{55}$}
\begin{minipage}[t]{14cm}
{\it Department of Particle Physics and Astrophysics, Weizmann
Institute, Rehovot, Israel}

\end{minipage}\\
\makebox[3em]{$^{56}$}
\begin{minipage}[t]{14cm}
{\it Department of Physics, University of Wisconsin, Madison,
Wisconsin 53706, USA}~$^{A}$

\end{minipage}\\
\makebox[3em]{$^{57}$}
\begin{minipage}[t]{14cm}
{\it Department of Physics, York University, Ontario, Canada M3J 1P3}~$^{N}$

\end{minipage}\\
\vspace{30em} \pagebreak[4]


\makebox[3ex]{$^{ A}$}
\begin{minipage}[t]{14cm}
 supported by the US Department of Energy\
\end{minipage}\\
\makebox[3ex]{$^{ B}$}
\begin{minipage}[t]{14cm}
 supported by the Italian National Institute for Nuclear Physics (INFN) \
\end{minipage}\\
\makebox[3ex]{$^{ C}$}
\begin{minipage}[t]{14cm}
 supported by the German Federal Ministry for Education and Research (BMBF), under
 contract No. 05 H09PDF\
\end{minipage}\\
\makebox[3ex]{$^{ D}$}
\begin{minipage}[t]{14cm}
 supported by the Science and Technology Facilities Council, UK\
\end{minipage}\\
\makebox[3ex]{$^{ E}$}
\begin{minipage}[t]{14cm}
 supported by HIR and UMRG grants from Universiti Malaya, and an ERGS grant from the
 Malaysian Ministry for Higher Education\
\end{minipage}\\
\makebox[3ex]{$^{ F}$}
\begin{minipage}[t]{14cm}
 supported by the US National Science Foundation. Any opinion,
 findings and conclusions or recommendations expressed in this material
 are those of the authors and do not necessarily reflect the views of the
 National Science Foundation.\
\end{minipage}\\
\makebox[3ex]{$^{ G}$}
\begin{minipage}[t]{14cm}
 supported by the Polish Ministry of Science and Higher Education as a scientific project No.
 DPN/N188/DESY/2009\
\end{minipage}\\
\makebox[3ex]{$^{ H}$}
\begin{minipage}[t]{14cm}
 supported by the Polish Ministry of Science and Higher Education and its grants
 for Scientific Research\
\end{minipage}\\
\makebox[3ex]{$^{ I}$}
\begin{minipage}[t]{14cm}
 supported by the German Federal Ministry for Education and Research (BMBF), under
 contract No. 05h09GUF, and the SFB 676 of the Deutsche Forschungsgemeinschaft (DFG) \
\end{minipage}\\
\makebox[3ex]{$^{ J}$}
\begin{minipage}[t]{14cm}
 supported by the Japanese Ministry of Education, Culture, Sports, Science and Technology
 (MEXT) and its grants for Scientific Research\
\end{minipage}\\
\makebox[3ex]{$^{ K}$}
\begin{minipage}[t]{14cm}
 supported by the Korean Ministry of Education and Korea Science and Engineering
 Foundation\
\end{minipage}\\
\makebox[3ex]{$^{ L}$}
\begin{minipage}[t]{14cm}
 supported by FNRS and its associated funds (IISN and FRIA) and by an Inter-University
 Attraction Poles Programme subsidised by the Belgian Federal Science Policy Office\
\end{minipage}\\
\makebox[3ex]{$^{ M}$}
\begin{minipage}[t]{14cm}
 supported by the Spanish Ministry of Education and Science through funds provided by
 CICYT\
\end{minipage}\\
\makebox[3ex]{$^{ N}$}
\begin{minipage}[t]{14cm}
 supported by the Natural Sciences and Engineering Research Council of Canada (NSERC) \
\end{minipage}\\
\makebox[3ex]{$^{ O}$}
\begin{minipage}[t]{14cm}
 partially supported by the German Federal Ministry for Education and Research (BMBF)\
\end{minipage}\\
\makebox[3ex]{$^{ P}$}
\begin{minipage}[t]{14cm}
 supported by RF Presidential grant N 3920.2012.2 for the Leading Scientific Schools and by
 the Russian Ministry of Education and Science through its grant for Scientific Research on
 High Energy Physics\
\end{minipage}\\
\makebox[3ex]{$^{ Q}$}
\begin{minipage}[t]{14cm}
 supported by the Netherlands Foundation for Research on Matter (FOM)\
\end{minipage}\\
\makebox[3ex]{$^{ R}$}
\begin{minipage}[t]{14cm}
 supported by the Israel Science Foundation\
\end{minipage}\\
\vspace{30em} \pagebreak[4]


\makebox[3ex]{$^{ a}$}
\begin{minipage}[t]{14cm}
now at University of Salerno, Italy\
\end{minipage}\\
\makebox[3ex]{$^{ b}$}
\begin{minipage}[t]{14cm}
now at Queen Mary University of London, United Kingdom\
\end{minipage}\\
\makebox[3ex]{$^{ c}$}
\begin{minipage}[t]{14cm}
also funded by Max Planck Institute for Physics, Munich, Germany\
\end{minipage}\\
\makebox[3ex]{$^{ d}$}
\begin{minipage}[t]{14cm}
also Senior Alexander von Humboldt Research Fellow at Hamburg University,
 Institute of Experimental Physics, Hamburg, Germany\
\end{minipage}\\
\makebox[3ex]{$^{ e}$}
\begin{minipage}[t]{14cm}
also at Cracow University of Technology, Faculty of Physics,
 Mathemathics and Applied Computer Science, Poland\
\end{minipage}\\
\makebox[3ex]{$^{ f}$}
\begin{minipage}[t]{14cm}
supported by the research grant No. 1 P03B 04529 (2005-2008)\
\end{minipage}\\
\makebox[3ex]{$^{ g}$}
\begin{minipage}[t]{14cm}
partially supported by the Polish National Science Centre projects DEC-2011/01/B/ST2/03643
 and DEC-2011/03/B/ST2/00220\
\end{minipage}\\
\makebox[3ex]{$^{ h}$}
\begin{minipage}[t]{14cm}
now at Rockefeller University, New York, NY
 10065, USA\
\end{minipage}\\
\makebox[3ex]{$^{ i}$}
\begin{minipage}[t]{14cm}
now at DESY group FS-CFEL-1\
\end{minipage}\\
\makebox[3ex]{$^{ j}$}
\begin{minipage}[t]{14cm}
now at Institute of High Energy Physics, Beijing, China\
\end{minipage}\\
\makebox[3ex]{$^{ k}$}
\begin{minipage}[t]{14cm}
now at DESY group FEB, Hamburg, Germany\
\end{minipage}\\
\makebox[3ex]{$^{ l}$}
\begin{minipage}[t]{14cm}
also at Moscow State University, Russia\
\end{minipage}\\
\makebox[3ex]{$^{ m}$}
\begin{minipage}[t]{14cm}
now at University of Liverpool, United Kingdom\
\end{minipage}\\
\makebox[3ex]{$^{ n}$}
\begin{minipage}[t]{14cm}
now at CERN, Geneva, Switzerland\
\end{minipage}\\
\makebox[3ex]{$^{ o}$}
\begin{minipage}[t]{14cm}
also affiliated with Universtiy College London, UK\
\end{minipage}\\
\makebox[3ex]{$^{ p}$}
\begin{minipage}[t]{14cm}
now at Goldman Sachs, London, UK\
\end{minipage}\\
\makebox[3ex]{$^{ q}$}
\begin{minipage}[t]{14cm}
also at Institute of Theoretical and Experimental Physics, Moscow, Russia\
\end{minipage}\\
\makebox[3ex]{$^{ r}$}
\begin{minipage}[t]{14cm}
also at FPACS, AGH-UST, Cracow, Poland\
\end{minipage}\\
\makebox[3ex]{$^{ s}$}
\begin{minipage}[t]{14cm}
partially supported by Warsaw University, Poland\
\end{minipage}\\
\makebox[3ex]{$^{ t}$}
\begin{minipage}[t]{14cm}
supported by the Alexander von Humboldt Foundation\
\end{minipage}\\
\makebox[3ex]{$^{ u}$}
\begin{minipage}[t]{14cm}
now at Istituto Nucleare di Fisica Nazionale (INFN), Pisa, Italy\
\end{minipage}\\
\makebox[3ex]{$^{ v}$}
\begin{minipage}[t]{14cm}
now at Haase Energie Technik AG, Neum\"unster, Germany\
\end{minipage}\\
\makebox[3ex]{$^{ w}$}
\begin{minipage}[t]{14cm}
now at Department of Physics, University of Bonn, Germany\
\end{minipage}\\
\makebox[3ex]{$^{ x}$}
\begin{minipage}[t]{14cm}
also affiliated with DESY, Germany\
\end{minipage}\\
\makebox[3ex]{$^{ y}$}
\begin{minipage}[t]{14cm}
also at University of Tokyo, Japan\
\end{minipage}\\
\makebox[3ex]{$^{ z}$}
\begin{minipage}[t]{14cm}
now at Kobe University, Japan\
\end{minipage}\\
\makebox[3ex]{$^{\dagger}$}
\begin{minipage}[t]{14cm}
 deceased \
\end{minipage}\\
\makebox[3ex]{$^{aa}$}
\begin{minipage}[t]{14cm}
supported by DESY, Germany\
\end{minipage}\\
\makebox[3ex]{$^{ab}$}
\begin{minipage}[t]{14cm}
member of National Technical University of Ukraine, Kyiv Polytechnic Institute,
 Kyiv, Ukraine\
\end{minipage}\\
\makebox[3ex]{$^{ac}$}
\begin{minipage}[t]{14cm}
member of National University of Kyiv - Mohyla Academy, Kyiv, Ukraine\
\end{minipage}\\
\makebox[3ex]{$^{ad}$}
\begin{minipage}[t]{14cm}
Alexander von Humboldt Professor; also at DESY and University of Oxford\
\end{minipage}\\
\makebox[3ex]{$^{ae}$}
\begin{minipage}[t]{14cm}
STFC Advanced Fellow\
\end{minipage}\\
\makebox[3ex]{$^{af}$}
\begin{minipage}[t]{14cm}
now at LNF, Frascati, Italy\
\end{minipage}\\
\makebox[3ex]{$^{ag}$}
\begin{minipage}[t]{14cm}
This material was based on work supported by the
 National Science Foundation, while working at the Foundation.\
\end{minipage}\\
\makebox[3ex]{$^{ah}$}
\begin{minipage}[t]{14cm}
also at Max Planck Institute for Physics, Munich, Germany, External Scientific Member\
\end{minipage}\\
\makebox[3ex]{$^{ai}$}
\begin{minipage}[t]{14cm}
now at Tokyo Metropolitan University, Japan\
\end{minipage}\\
\makebox[3ex]{$^{aj}$}
\begin{minipage}[t]{14cm}
now at Nihon Institute of Medical Science, Japan\
\end{minipage}\\
\makebox[3ex]{$^{ak}$}
\begin{minipage}[t]{14cm}
now at Osaka University, Osaka, Japan\
\end{minipage}\\
\makebox[3ex]{$^{al}$}
\begin{minipage}[t]{14cm}
also at \L\'{o}d\'{z} University, Poland\
\end{minipage}\\
\makebox[3ex]{$^{am}$}
\begin{minipage}[t]{14cm}
member of \L\'{o}d\'{z} University, Poland\
\end{minipage}\\
\makebox[3ex]{$^{an}$}
\begin{minipage}[t]{14cm}
now at Department of Physics, Stockholm University, Stockholm, Sweden\
\end{minipage}\\
\makebox[3ex]{$^{ao}$}
\begin{minipage}[t]{14cm}
also at Cardinal Stefan Wyszy\'nski University, Warsaw, Poland\
\end{minipage}\\

}


%% file: ijphp-txt.tex
\newcommand{\palong}{P$_{\mbox{\tiny{along }}}$}
\newcommand{\pagainst}{P$_{\mbox{\tiny{against }}}$}

\pagenumbering{arabic} 
\pagestyle{plain}
\section{Introduction}
\label{sec:int}

The inelastic production of  \(J/\psi\) and of \(\psi^{\prime}\)
has been studied 
for several years in 
hadron and electron-proton colliders and in fixed target 
experiments~\cite{epj:c71:1534}. 
At HERA, the reactions 
\begin{equation}
ep \to e J/\psi X,
\label{eq:jpsi}
\end{equation}
and
\begin{equation}
ep \to e \psi^\prime X,
\label{eq:psiprime}
\end{equation}
have been studied~\cite{epj:c27:173,epj:c68:401} 
for low virtuality of the exchanged photon 
(photoproduction) in the range $z < 0.9$, where $z$ denotes the 
fraction of the incident photon energy carried by the meson in 
the proton rest frame, thus excluding the diffractive process for 
which $z \sim 1$. 
In the HERA photoproduction regime, the production of inelastic 
$J/\psi$  or $\psi^\prime$ mesons arises mostly from direct and 
resolved photon interactions. In leading-order (LO) Quantum 
Chromodynamics (QCD), the two processes can be distinguished; in 
direct-photon processes the photon enters directly into the hard 
interaction; in resolved-photon processes the photon acts as a source 
of partons, one of which participates in the hard interaction.
The inelastic process in the photoproduction region is dominated by 
photon--gluon fusion.
In this direct-photon process the photon emitted from the incoming 
electron interacts with a gluon from the proton to produce a pair of 
charm-anticharm quarks, $c\bar{c}$, which then turn into the $J/\psi$ 
or the $\psi^\prime$ mesons. When the $c\bar{c}$ pair emerges from the
hard process with the 
quantum numbers of the mesons, the reaction is described in the framework 
of perturbative Quantum Chromodynamics (pQCD) by models such as the 
Colour Singlet (CS) model. In the Colour Octet (CO) model, the 
$c\bar{c}$ pair emerges from the hard process with quantum numbers 
different from those of the mesons and emits one or more soft gluons before 
turning into the physical meson state. 
Examples of direct-photon LO diagrams with a CS and a CO hard subprocess 
are shown in Fig.~\ref{fig:fey}.

Full next-to-leading order (NLO) $J/\psi$ cross section predictions 
using only the direct-photon CS contributions have already been
performed~\cite{pl:b348:657,*np:b459:3,pr:d80:34020}.
The non-relativistic QCD framework (NRQCD)~\cite{pr:d51:1125} allows 
the evaluation of $J/\psi$ cross sections including direct and resolved 
photon processes with CS and CO contributions. The former contribution
can be thought of as the first term of the NRQCD expansion and so it is
an integral component of this theoretical formalism.
Recently, the full computation was performed in the HERA photoproduction
regime at the NLO level~\cite{prl:104:072001,*pr:d84:051501}.
The numerical values of the CS and CO matrix elements were obtained from
a global fit to hadroproduction, electroproduction and photoproduction
inelastic $J/\psi$ data~\cite{prl:104:072001,*pr:d84:051501}.

$J/\psi$ cross sections have also been 
evaluated~\cite{epj:c27:87,epj:c71:1631} in the \(k_T\)--factorisation 
approach~\cite{prep:100:1,*sovjnp:53:657,*np:b366:135,*np:b360:3}.
In this model, based on non--collinear parton dynamics governed by the
CCFM~\cite{np:b296:49,*pl:b234:339} evolution equations,
effects of non--zero gluon transverse momentum are taken into account. 
Cross sections are then calculated as the convolution of unintegrated,
transverse--momentum dependent gluon densities and LO off--shell matrix 
elements. Direct and resolved photon processes are included. The matrix 
elements are computed in the CS model.

Measurements of the reactions (\ref{eq:jpsi}) and (\ref{eq:psiprime}) 
have been previously performed by the ZEUS collaboration~\cite{epj:c27:173}, 
using an integrated luminosity of 38 pb$^{-1}$, and by the H1 
collaboration~\cite{epj:c68:401}, using an integrated luminosity of 
165 pb$^{-1}$. 
Total and differential cross sections were presented as a function of 
various kinematical variables. 
The H1 and ZEUS collaborations have also published a measurement of the 
$J/\psi$ helicity distribution~\cite{epj:c68:401,jhep12:2009:007}, the
ZEUS result was obtained using the full HERA luminosity.
LO and NLO QCD predictions, as well as LO NRQCD calculations, were compared 
to the measurements. 
None of the 
calculations could describe the data in the whole kinematic range of 
the measurements. The data were shown to have the potential to reduce 
the large uncertainties in the phenomenological parameters used
in the calculations.

In this paper, measurements of reactions (\ref{eq:jpsi}) 
and (\ref{eq:psiprime}) are presented using a luminosity of 468 pb$^{-1}$.
The \(J/\psi\) and \(\psi^{\prime}\) mesons were identified using the 
$\mu^+ \mu^-$ decay modes. 

The \(\psi^{\prime}\) to \(J/\psi\) cross section ratio was measured in 
the range 60 $ < W < $ 190 GeV and $0.55 < z < 0.9$ as a function of
$W$, $z$ and $p_T$.
Here $W$ is the $\gamma p$ centre-of-mass energy and $p_T$ 
is the transverse momentum of the mesons with respect to the beam axis.
The cross sections for inelastic $J/\psi$ photoproduction as a function 
of $p^2_T$, for different $z$ ranges, and as a function of $z$, for 
different $p_T$ ranges, were measured in the range 60 $ < W < $ 240 GeV,
$0.1 < z < 0.9$ and $p_T > 1$ GeV.
The momentum flow along and against the \(J/\psi\) direction of flight in 
the laboratory frame, as obtained from the charged tracks produced together 
with the \(J/\psi\) in the range 60 $ < W < $ 240 GeV, $0.3 < z < 0.9$ and 
$1 < p_T < 10$ GeV, was studied in order to shed further light on 
the production mechanisms. 

\section{Experimental set-up}
\label{sec:setup}

The analysis presented here is based on data collected by the ZEUS 
detector at HERA in the period 1996--2007. 
In 1998--2007 (1996--1997), HERA provided electron\footnote{Here and 
in the following, 
the term ``electron'' denotes 
generically both the electron ($e^-$) and the positron ($e^+$).}
beams of energy $E_e$~=~27.5 GeV and proton beams of energy $E_p = 920 
~(820)$ GeV, resulting in a centre-of-mass energy of $\sqrt{s} = 318 ~(300)$ 
GeV, giving an integrated luminosity of $430~(38)$ pb$^{-1}$.

A detailed description of the ZEUS detector can be found
elsewhere~\cite{pl:b293:465,zeus:1993:bluebook}. A brief outline of the 
components that are most relevant for this analysis is given below.

Charged particles were tracked in the central tracking detector 
(CTD)~\citeCTD, which operated in a magnetic field of 
$1.43\Tesla$ provided by a thin superconducting coil. 
Before the 2003--2007 running period, the ZEUS tracking system was 
upgraded with a silicon microvertex detector (MVD)~\cite{nim:a581:656}. 
In the following, the term ``CTD-MVD track'' denotes
generically both the tracks measured in the CTD and (after 2002)
in the CTD and MVD.

The high-resolution uranium--scintillator calorimeter (CAL)~\citeCAL 
consisted of three parts: the forward (FCAL), the barrel (BCAL) and the rear 
(RCAL) calorimeters\footnote{The ZEUS coordinate system is a 
right-handed Cartesian system, with the $Z$ axis pointing in the 
proton--beam direction, referred to as the ``forward direction'', and the $X$ 
axis pointing towards the centre of HERA.
The coordinate origin is at the nominal interaction point. 
The polar angle, $\theta$, is measured with respect to the proton--beam 
direction. The pseudorapidity is defined as
$\eta$=--ln(tan $\frac{\theta}{2}$).}. 
Each part was subdivided transversely into towers and longitudinally into 
one electromagnetic section (EMC) and either one (in RCAL) or two (in BCAL 
and FCAL) hadronic sections (HAC). 
The smallest subdivision of the calorimeter was called a cell.  
The CAL energy resolutions, as measured under test-beam conditions, were 
$\sigma(E)/E=0.18/\sqrt{E}$ for electrons and
$\sigma(E)/E=0.35/\sqrt{E}$ for hadrons ($E$ in $\Gev$).
The timing resolution of the CAL was better than 1 ns for energy deposits 
greater than 4.5 GeV.

Muons were identified as tracks measured in the barrel and rear 
muon chambers (BMUON and RMUON)~\cite{nim:a333:342}.
The muon chambers were placed inside and outside the magnetised iron 
yoke surrounding the CAL.
The barrel and rear inner muon chambers (BMUI and RMUI) covered the 
polar-angle regions  $34^\circ < \theta < 
 135^\circ$ and $135^\circ < \theta < 171^\circ$, respectively.

The luminosity was measured using the Bethe--Heitler reaction 
$ep \rightarrow e \gamma p$ with the luminosity detector 
which consisted of a lead--scintillator calorimeter 
\cite{desy-92-066,*zfp:c63:391,*acpp:b32:2025} and, after 2002, 
of an additional magnetic spectrometer\cite{nim:a565:572} system.
The fractional systematic uncertainty on the measured luminosity was
1.9\%.

\section{Event selection and kinematic variables}
\label{sec:event}

The online and offline selections, as well as the reconstruction of the 
kinematic variables, closely follow the previous 
analysis~\cite{epj:c27:173}.

Online, the BMUI and RMUI chambers were used to tag muons by matching 
segments in the muon chambers with CTD-MVD tracks, as well as 
with energy deposits in the CAL consistent with the passage of a 
minimum-ionising particle (m.i.p.).

The different steps of the offline selection procedure are
described in the following paragraphs.
An event was accepted if it had two primary-vertex CTD-MVD tracks with 
invariant mass between 2 -- 5 GeV.
One track had to be identified in the inner muon chambers and 
matched to a m.i.p. cluster in the CAL. It was required to have a momentum 
greater than 1.8 GeV if it was in the rear region or a transverse 
momentum greater than 1.4 GeV if in the barrel region. The other track 
had to be matched to a m.i.p. cluster in the CAL and was required 
to have a transverse momentum greater than 0.9 GeV. 
Both tracks were restricted to the 
pseudorapidity region $|\eta |<$ 1.75.
To reject cosmic rays, events in which the angle between the two muon 
tracks was larger than 174$^{\circ}$ were removed. 

In addition, events were required to have a calorimetric
energy deposit larger than 1 
GeV in a cone of 35$^{\circ}$ around the forward direction (excluding 
possible calorimeter deposits due to the decay muons). 
This requirement completely rejects exclusively produced \(J/\psi\) mesons, 
\(e p \rightarrow e p J/\psi\).
It also strongly suppresses the background from proton 
diffractive--dissociation, \(e p \rightarrow e N J/\psi\), because the 
low invariant
mass hadronic system $N$ can often (but not always) escape along the outgoing 
proton direction without any activity in the FCAL.
A reduction of the remaining background is achieved by requiring the 
events to have, in addition to the two decay muon tracks, at 
least one additional track with transverse momentum larger than 250 MeV 
and pseudorapidity $|\eta |<$ 1.75.

The \(\psi^{\prime}\) production in proton diffractive--dissociation 
processes with the decay chain
\(J/\psi (\rightarrow \mu^+ \mu^-)\) \(\pi^+\) \(\pi^-\)
was identified in the selected data sample. For the bulk of these
events only four charged tracks are visible in the detector.
For the events with a \,\(\mu^+ \mu^-\) invariant mass, $m_{\mu \mu}$, 
in the interval [2.85, 3.30] \(\gev\), and with exactly two additional 
primary--vertex tracks of opposite charge, the total invariant mass $m_4$ 
of the four tracks was evaluated.
Events with a mass difference $m_4-m_{\mu \mu}$
within $\pm 60\;$MeV of the nominal mass difference 
$m_{\psi^{\prime}} - m_{J/\psi} = 589\;$MeV~\cite{jp:g37:075021} 
were discarded.
This topology was tagged only in 1.2\% of the overall selected \(J/\psi\) 
sample and removed. 

These requirements effectively select inelastic \(J/\psi\) and 
\(\psi^{\prime}\) mesons.
\(J/\psi\) and  \(\psi^{\prime}\) mesons from decays of $b$ hadrons 
are also included in the data sample. 

The kinematic region considered was defined by the inelasticity 
variable $z$ and by the photon-proton centre-of-mass energy
\begin{equation}
\label{eq:W_def}
W^2 = (P +q)^2 ,
\end{equation}
where $P$ and $q$ are the four--momenta of the incoming 
proton and the exchanged photon, respectively. 
It was calculated using   
\begin{equation}
\label{eq:W_comp}
W^2 = 2 E_p (E-p_Z) ,
\end{equation}
where $(E-p_Z)$, the difference between the energy and the momentum
along the $Z$ axis, is summed over all final-state energy-flow
objects~\cite{epj:c1:81,*thesis:briskin:1998}
(EFOs) which combine the information from calorimetry and tracking.

The inelasticity $z=  \frac{P \cdot p_{\psi} }{P \cdot q}$ 
was determined as  
\begin{equation}
\label{eq:z_def}
z = \frac{(E-p_Z)_{\psi}}{(E-p_Z)},
\end{equation}
where $\psi$ can be either a \(J/\psi\) or a \(\psi^{\prime}\) meson,
$p_{\psi}$ is the four-momentum of the \(\psi\) and 
$(E-p_Z)_{\psi}$ was calculated using the two tracks forming the 
\(\psi\).

In order to reject deep inelastic scattering, events
were required to have $E-p_Z < 32$ \(\gev\).
This restricts the virtuality of the exchanged photon, $Q^2 = -q^2$,
to $Q^2 \lesssim 1$ \(\gev^2\), with a median of about 10$^{-4}$ 
\(\gev^2\). 
The elimination of deep inelastic scattering events was independently 
confirmed by searching for scattered electrons in the 
CAL~\cite{nim:a365:508}; none was found.

Table~\ref{tab:kinrange} summarises the various kinematic regions used
for the presented measurements.

\section{Monte Carlo models}
\label{sec:mcs}

The inelastic production of \(J/\psi\) and \(\psi^{\prime}\) mesons 
was simulated using the {\sc Herwig} 6.100~\cite{cpc:67:465}
program, which generates direct photon events according to the 
LO diagrams of the photon--gluon fusion process,
\(\gamma g \rightarrow \psi g\).
The processes are calculated in the framework of the CS model. 
The {\sc Herwig} MC provides in general a good description of the data.
To improve the agreement further, the $p_T$ spectrum 
was reweighted to the data. The average weight of the MC events with
$p_T$ around 1 \(\gev\) is 0.85. The average weight for $p_T > 4$ \(\gev\) 
is instead 1.8.

Diffractive production of \(J/\psi\) and \(\psi^{\prime}\) mesons with 
proton dissociation was simulated with the 
{\sc Epsoft}~\cite{thesis:kasprzak:1994} MC generator, which was tuned to 
describe such processes at HERA~\cite{thesis:adamczyk:1999}.

The {\sc Pythia~6.220} MC 
generator~\cite{cpc:135:238,*epj:c17:137,*hep-ph-0108264}
was used to generate \(J/\psi\) and \(\chi_c\) states from the 
resolved-photon process, with LO matrix elements computed in the CS model.
The generator cross sections for the \(J/\psi\) and \(\chi_c\) states are 
very similar.
For the generation of the $\chi_{c1}(1P)$ and $\chi_{c2}(1P)$
mesons, only the \(J/\psi\) \(\gamma\) decay channel was considered. 
The final state photon is at low energy, $O(400)$\(\mev\), 
basically indistinguishable from the remaining hadronic activity of the 
event.
Hence the effective resolved-photon \(J/\psi\) contribution can be thought
of as due to the genuine resolved-photon component plus the $\chi_{c}$
feed--down.
The resolved \(\psi^{\prime}\) contribution was neglected due to the
small resolved-to-direct cross section ratio and to the additional
reduction due to the $\psi^{\prime} \rightarrow J/\psi X$ branching ratio.

The {\sc Pythia} MC was also used to generate the production
of \(J/\psi\) and \(\psi^{\prime}\) mesons originating from $b$ hadron 
decays, mostly from $B$-mesons.
The following beauty-quark production processes were generated
(according to the {\sc Pythia} notation):
direct, resolved, $\gamma$ and proton excitation. 
The beauty-quark mass was set to 4.75 GeV and the branching ratios of 
the $b$ hadrons to \(J/\psi\) and \(\psi^{\prime}\) were set to the 
corresponding PDG~\cite{jp:g37:075021} values.

All generated events were passed through a full simulation of the ZEUS 
detector based on {\sc Geant} 3~\cite{tech:cern-dd-ee-84-1}. 
They were then subjected to the same trigger requirements and processed 
by the same reconstruction program as the data.

\section{Signal determination and cross sections calculation}
\label{sec:cross-sections}

The invariant-mass spectrum of the muon pairs measured in the phase 
space region used in the determination of the \(\psi^{\prime}\) 
to \(J/\psi\) cross section ratio,
$60 < W < 190$ \(\gev\) and $0.55 < z < 0.9$, is shown in 
Fig.~\ref{fig:mmumu-pap}.
A non-resonant background contribution, mostly due to hadrons 
misidentified as muons, is also visible.
This contribution was estimated by fitting the product of a second-order
polynomial and an exponential function to the region 2--2.75 and 3.8--5
\(\gev\), outside the \(J/\psi\) and \(\psi^{\prime}\) invariant-mass 
window.
The number of \(J/\psi\) events was obtained by subtracting the number of 
background events, estimated from the fit procedure, from the total number
of events inside the \(J/\psi\) invariant-mass window, 2.85--3.3 GeV. 
This procedure resulted in 11295 $\pm$ 114 \(J/\psi\) events.
The same procedure applied to the \(\psi^{\prime}\) invariant-mass
window, 3.55--3.8 GeV, gave 448 $\pm$ 34 events.

Applying the same procedure to the phase space region used for the 
differential $J/\psi$ cross section measurements, $60 < W < 190$ \(\gev\), 
$0.1 < z < 0.9$ and $p_T > 1$ \(\gev\), 12671 $\pm$ 161 \(J/\psi\) events 
were found. The fitting procedure described above was performed for each
measurement bin presented in this paper.

The cross section for any observable, ${\mathcal O}$, was computed 
for each bin, $i$, using correction factors, $C_i({\mathcal O})$,  
defined as
$C_i({\mathcal O})=N_i^{\rm gen}({\mathcal O})/
                   N_i^{\rm rec}({\mathcal O})$,
where $N_i^{\rm gen}({\mathcal O})$ is the number of events generated 
with the {\sc Herwig} MC and $N_i^{\rm rec}({\mathcal O})$ is the number 
of the events reconstructed by the standard analysis chain.
The factors $C_i({\mathcal O})$ take into account the overall acceptance
including the geometrical acceptance and the detector, trigger
and reconstruction efficiencies.
They also take into account bin-to-bin migrations.

For $0.9<z<1$, the events are largely diffractive.
Therefore, the analysis of inelastic $J/\psi$ production was restricted 
to the region $0.1 < z < 0.9$. 
In order to further suppress diffractive events, the transverse momentum 
of the $J/\psi$ mesons had to fulfill $p_T > 1$ \(\gev\).
The remaining contamination was estimated by fitting the relative fractions 
of non-diffractive and diffractive events to the data $z$-distribution,
using the {\sc Herwig} and {\sc Epsoft} MC simulations as templates.
From this fit, the overall diffractive background contribution 
for $0.1 < z < 0.9$ is 4.6 $\pm$ 1.6\%. 

In Fig.~\ref{fig:ccpl-pap} the {\sc Herwig} and {\sc Epsoft} MC mixture,
in the kinematic region 60 $ < W < $ 240 \(\gev\), $0.3 < z < 0.9$ and 
$p_T > 1$ \(\gev\), is compared to the data: a reasonable description
is found. The region $0.1 < z < 0.3$ was removed because not diffractive
background is present at low $z$.
The estimated diffractive background was subtracted bin by bin from 
the measured differential cross sections.

The cross sections measured in this analysis include also
contributions from resolved-photon processes and from 
decays of beauty hadrons.
Inelastic \(J/\psi\) production via the resolved-photon process has not 
been measured explicitly up to now in the photoproduction regime.
QCD predictions, as well as the {\sc Pythia} MC simulation described
in Section~\ref{sec:mcs}, indicate that this contribution is largest
at low $z$ values. For $z < 0.1$, the expected size of this 
contribution can be larger than the direct-photon component.
However, for $z > 0.1$, the resolved-to-direct photon production 
ratio is expected to be small. Since the acceptances obtained from 
the {\sc Herwig} and {\sc Pythia} MC simulations are similar,
the {\sc Herwig} MC alone was used for the overall acceptance
corrections.

The contribution to the measured cross sections due to \(J/\psi\)
originating from $B$ meson decays was estimated using the inclusive
beauty {\sc Pythia} MC sample described in Section~\ref{sec:mcs}.
The simulation predictions were scaled by a factor 1.11
according to the recent ZEUS measurement~\cite{epj:c71:1659}
of beauty photoproduction\footnote{The scaling factors obtained
in the measurements~\cite{epj:c71:1659,jhep02:2009:032} vary between 1.11 
and 1.84}.
This leads to the estimation that on average 1.6\% of the observed \(J/\psi\) 
mesons originated from beauty hadron decays.
The largest relative contribution, 4.5\%, is in the kinematic
region $0.1 < z < 0.3$ and $1 < p_T^2 <2$ \(\gev^2\).
This component is not subtracted from the measured cross sections.

\section{Systematic uncertainties}

For all the measured quantities,
the following sources of systematic uncertainties were investigated
(their effects on the measured cross sections are given in parentheses):

\begin{itemize}

\item muon trigger and reconstruction efficiencies: the BMUI and RMUI 
muon chamber 
efficiencies were extracted from the data using muon pairs from elastic 
$J/\psi$ events and from the process $\gamma \gamma \rightarrow \mu^+ \mu^-$.
These efficiencies take into account the full muon acquisition chain,
from the online to the offline level and are known with a $\pm 5\%$ 
uncertainty ($5\%$ uniformly distributed in $p_T$ and $z$);
 
\item hadronic energy resolution: the $W$ and $z$ resolutions are dominated 
by the hadronic energy resolution affecting the quantity $(E-p_Z)$.
The hadronic $(E-p_Z)$ resolution in the MC 
was smeared event by event by $\pm 20\%$, a conservative upper limit 
of a possible systematic difference between data and MC.
This gave only small cross sections variations ($< 5 \%$);

\item {\sc Herwig} MC $p_T$ spectrum: the $p_T$ spectrum of the $J/\psi$ 
mesons in the {\sc Herwig} MC simulation was varied within ranges allowed 
by the comparison between data and simulation and the correction factors 
were re-evaluated ($< 5 \%$);

\item $J/\psi$ helicity distribution: the $J/\psi$ helicity distribution
can be described by two parameters $\lambda$ and
$\nu$~\cite{prl:107:232001}. In the {\sc Herwig} MC these are set to zero. 
According to the direct measurement of the helicity parameters performed by 
ZEUS~\cite{jhep12:2009:007},
all data points lie within the region of the $\lambda$-$\nu$ plane
defined by $| \lambda | <$ 0.5 and $| \nu | <$ 0.5 with only a mild
$p_T$ or $z$ dependence.
Hence, as a systematic check, the {\sc Herwig} MC was reweighted varying 
independently $\lambda$ and $\nu$ in the range $\pm 0.5$
and the correction factors were re-evaluated
($5-10\%$ depending on the $p_T$ and $z$ region);

\item diffractive simulation: the {\sc Epsoft} MC simulation parameters were 
varied within ranges allowed by the comparison between data and the 
{\sc Epsoft} MC simulation in the region $0.9<z<1$. 
The diffractive background was re-evaluated ($< 5 \%$ at high $z$ and low 
$p_T$, negligible elsewhere);

\item diffractive subtraction: the relative fraction of inelastic and
diffractive processes, as represented by the {\sc Herwig} and 
{\sc Epsoft} MC, was 
fixed by the procedure described in Section~\ref{sec:cross-sections}.
It is known to a precision limited by the number of $J/\psi$ events in 
the data and the process modeling by the MCs.
The relative fractions were varied within ranges allowed by the comparison 
between data and simulation (up to $10 \%$ at high $z$ and low $p_T$, 
negligible elsewhere);

\item invariant--mass window: the $m_{\mu^{+}\mu^{-}}$ invariant--mass window 
used to estimate the number of \(J/\psi\) events above the non-resonant 
background was enlarged to $[2.8,3.35]$ GeV and tightened to $[2.9,3.3]$ GeV.
For the \(\psi^{\prime}\) to \(J/\psi\) cross section ratios, 
similar mass window variations were also applied for the \(\psi^{\prime}\) 
signal (generally $< 5 \%$, up to $10 \%$ at low $z$ values where the number 
of expected and observed events is small and the non-resonant background 
is largest);

\item additional track cut: the requirement of three tracks, including
the two \(J/\psi\) decay muons, with transverse momentum 
larger than 250 MeV and pseudorapidity $|\eta |<$ 1.75, was replaced by 
the requirement of five tracks with transverse momentum larger than 
125 MeV, in the same pseudorapidity range. 
With this stronger requirement the diffractive \(J/\psi\)
background and the diffractive \(\psi^{\prime}\) contribution via the 
cascade decay \(J/\psi (\rightarrow \mu^+ \mu^-)\) \(\pi^+\) \(\pi^-\) 
are expected to vanish.
Furthermore, a change in the overall multiplicity cut allows a test of how
well the MC model reproduces the data in this respect.
The MC mixture gives a fair description of the track multiplicity observed 
in the data.
The cross sections were re-evaluated with the harder multiplicity cut
(generally $< 5 \%$, up to $20 \%$ in some bins at low $z$ and high $z$
high $p_T$).

\end{itemize}

All of the above individual sources of systematic uncertainty were 
added in quadrature.

The following sources would result in an overall small shift of the cross 
sections:
\begin{itemize}
\item the integrated luminosity determination gave an uncertainty of 
$\pm 1.9 \%$;
\item the $J/\psi \rightarrow \mu^+\mu^-$ branching ratio,
$5.93 \pm 0.06 \%$~\cite{jp:g37:075021}, gave an uncertainty of $\pm 1 \%$.
\end{itemize}
They were not included. 

\section{Results}
\label{sec:res}

\subsection{\(\psi^{\prime}\) to \(J/\psi\) cross section ratio}
\label{subsec:2sto1s}

The \(\psi^{\prime}\) to \(J/\psi\) cross section ratio 
was measured using the rates of \(\psi^{\prime} \rightarrow \mu^+\mu^-\)
and \(J/\psi \rightarrow \mu^+\mu^-\). The 
ratio was determined in the 
region $60 < W < 190$ \(\gev\), 0.55 $< z <$ 0.9.
The $p_T > 1$ \(\gev\) requirement was removed to maximise the available 
statistics. An increase of the diffractive background is expected. But 
under the assumption that this background contribution will be the
same for the \(\psi^{\prime}\) and \(J/\psi\) mesons it will cancel in
the cross section ratio.
The range $190 < W < 240$ \(\gev\) and $0.1 < z < 0.55$ was not included
because the \(\psi^{\prime}\) peak was not visible in this high $W$
and low $z$ region.
The \(\psi^{\prime}\) to \(J/\psi\) cross section ratio was computed in bins 
of $W$, $z$ and $p_T$ from
\begin{eqnarray}
\frac{\sigma_i(\psi^\prime)}{\sigma_i(J/\psi)} = 
\frac{N_i^{2S}}{N_i^{1S}} \cdot
\frac{C_i^{1S}}{C_i^{2S}} \cdot
\frac{Br^{\mu}}{Br^{\mu^{\prime}}} \cdot 
\left(1-\frac{N_i^{2S}}{N_i^{1S}} ~
\frac{C_i^{1S}}{C_i^{2S}} ~
\frac{Br^{\mu}}{Br^{\mu^{\prime}}} ~
Br^{\prime} \right)^{-1} ,
\label{eq-sigrat}
\nonumber
\end{eqnarray}
where, for the considered bin $i$, $N^{1S}_i$ ($N^{2S}_i$) is the number 
of \(J/\psi\) (\(\psi^{\prime}\)) events observed, $C^{1S}_i$ ($C^{2S}_i$) 
is the correction factor (see Section~\ref{sec:cross-sections}) computed 
using the {\sc Herwig} MC, 
$Br^{\mu}$ ($Br^{\mu^{\prime}}$) is the $J/\psi$ ($\psi^{\prime}$) muonic 
branching ratio and $Br^{\prime}$ is the $\psi^{\prime} \rightarrow J/\psi~X$ 
branching ratio. The values used are $Br^{\mu} = 5.93 \%$, 
$Br^{\mu^{\prime}} = 0.77 \%$ and 
$Br^{\prime} = 59.5 \%$~\cite{jp:g37:075021}.
With this technique, the cross section ratio was corrected for the 
\(\psi^{\prime} \rightarrow J/\psi~(\rightarrow \mu^+\mu^-)~X\) cascade 
decay.

Since NLO predictions are not available for \(\psi^{\prime}\),
only the LO CS model expectations can be compared to the
data.
In the CS model, the underlying production mechanism is the same for
\(J/\psi\) and \(\psi^{\prime}\), hence 
all cross section ratios should be largely independent of the kinematic 
variables.
Using the values of $Br^{\mu}$ and $Br^{\mu^{\prime}}$ given above, the
expected ratio is 0.25~\cite{pl:b348:657,*np:b459:3}.
Since the NLO corrections, though being large, should be similar for
\(J/\psi\) and \(\psi^{\prime}\), the ratio at NLO is not 
expected to differ significantly from that at LO.

The results, shown in Fig.~\ref{fig:2sto1s} and listed in 
Table~\ref{tab:ratio}, are dominated by the statistical uncertainties while
most of the systematic uncertainties cancel in the ratio.
The LO CS predictions agree reasonably well with the data.

\subsection{\(J/\psi\) differential cross sections}
\label{subsec:crossect}

The \(J/\psi\) differential cross sections presented here include the 
inelastic \(\psi^{\prime}\) feed--down via the decay
$ \psi^\prime \rightarrow J/\psi~(\rightarrow \mu^+\mu^-)~X$ and the
contribution from $b$ hadron decays. The \(\psi^{\prime}\) feed--down
contributes about $15 \%$ and the $b$ hadron decays $1.6 \%$ (see 
Section~\ref{sec:cross-sections}).
The $W$ range of the differential cross sections is $60 < W < 240$ \(\gev\).

The differential cross sections $d\sigma/dp_T^2$ were measured 
in the range 1 $< p^2_T<$ 100 \(\gev^2\) for different $z$ ranges.
The results are listed in Table~\ref{tab:2diffpt} and shown in 
Figs.~\ref{fig:dsdpt2-nlocs+co} and~\ref{fig:dsdpt2-kt}.
The predictions of a NRQCD calculation~\cite{prl:104:072001,*pr:d84:051501} 
are compared to the data in Fig.~\ref{fig:dsdpt2-nlocs+co} and those
based on the $k_T$--factorization approach~\cite{epj:c71:1631} in
Fig.~\ref{fig:dsdpt2-kt}.\footnote{Both the NRQCD and the 
\(k_T\)--factorisation calculations do not include \(\psi^{\prime}\) 
feed--down and $b$ hadron decays, however these expected contributions are 
small compared to the uncertainties of the calculations.}

The differential cross sections $d\sigma/dz$ were measured in the range 
0.1 $< z <$ 0.9 for different $p_T$ ranges. 
The results are shown in Figs.~\ref{fig:dsdz-nlocs+co} and~\ref{fig:dsdz-kt} 
and listed in Table~\ref{tab:2diffz}.

The present measurements are in agreement with the results obtained by the 
H1 collaboration~\cite{epj:c68:401} except in the region $z > 0.6$ and
$p_T > 3$ \(\gev\) where the ZEUS cross sections are above the H1 
measurements.

\subsubsection{Comparison of NRQCD calculation}

In Fig.~\ref{fig:dsdpt2-nlocs+co} a 
prediction~\cite{prl:104:072001,*pr:d84:051501} performed in the NRQCD 
framework including direct and resolved photon processes is compared
to the measured $d\sigma/dp_T^2$.
The hard subprocesses take into account both CS and CO terms to NLO.
The square of the renormalisation and factorisation scales used is
\(4 \cdot m_c^2 + p_T^2 \), the charm quark mass, $m_c$, is set to 
1.5 \( \gev \) and the strong coupling constant, 
\( \alpha_s(M_Z)\), to 0.118. 
The NRQCD scale, connected to the colour-octet terms, is set to $m_c$.
The CS contribution alone predicts cross sections significantly below the
data\footnote{
The NLO CS predictions~\cite{pl:b348:657,*np:b459:3} shown in the previous 
publication~\cite{epj:c27:173} were the first performed and used extreme
values for the renormalisation and factorisation scales, with the effect of 
artificially increasing the normalisation of the predicted 
cross sections~\cite{proc:eps-hep:2009:076}.}
and fails to describe the data in all $z$ regions shown here.
Including CO 
terms give a dramatic improvement and leads to a rough agreement
with the data. In general the calculation reproduces the steep drop of 
$d\sigma/dp_T^2$ with $p_T^2$, however, in the intermediate $z$ range,
$0.3 < z < 0.75$, the prediction rises less steeply than the data towards 
the smallest values of  $p_T^2$.

In Fig.~\ref{fig:dsdz-nlocs+co} the NRQCD predictions described above
are compared to the measured $d\sigma/dz$.
The predictions rise too steeply with $z$ compared to the data, for 
all the $p_T$ ranges.

\subsubsection{Comparison of \(k_T\)--factorisation approach}

In Fig.~\ref{fig:dsdpt2-kt} a prediction~\cite{epj:c71:1631} performed 
in the \(k_T\)--factorisation approach is compared to the measured
$d\sigma/dp_T^2$.
The matrix 
elements are computed in the CS model using \( m_c \) = 1.5 \(\gev\)
and \( \alpha_s(M_Z) = 0.1232 \).
In the numerical calculation, the renormalisation and factorisation scales
squared are set to \( m_{J/\psi}^2+p_T^2 \) and \( \hat{s} + Q_T^2 \),
respectively, where \(\hat{s}\) is the four-momentum squared of the hard 
subprocess
and \(Q_T \) is the transverse momentum of the initial parton.
The unintegrated CCFM parton density~\cite{ccfmj2003set1}
was selected.
Using different sets of parton densities leads to changes in the prediction
that are small with respect to the effects of scale variations already shown 
in Fig.~\ref{fig:dsdpt2-kt}. Thus this source of theoretical uncertainties
was neglected.
The \(k_T\)--factorisation prediction, with the values of $m_c$ and
$\alpha_s$ given above, provides a better description of the data
than the NRQCD model.

The above \(k_T\)--factorisation predictions are compared to the 
differential cross sections $d\sigma/dz$ in Fig.~\ref{fig:dsdz-kt}. Here 
too the description is better than that of the NRQCD model. Note however 
that the \(k_T\)--factorisation model prediction suffers from large 
theoretical uncertainties, in particular at low  $p_T$.

\subsection{Momentum flow along and against the \(J/\psi\) direction}

As pointed out by Brambilla et al.~\cite{epj:c71:1534}, the different 
colour flow in CS and CO hard subprocesses is expected to
translate into different properties of the hadronic final state.
In the photoproduction regime, the transverse momentum of the incoming
photon is negligible. Thus in the CS model (see Fig.~\ref{fig:fey} (a)),
at LO the \(J/\psi\) and the final state gluon are expected to be back to back.
Hence, in this model, the momentum flow along the \(J/\psi\) direction, 
\palong, is expected to be small. The momentum flow against the \(J/\psi\) 
direction, \pagainst, should instead be driven by the hadronisation of 
the gluon. In the CO framework (see Fig.~\ref{fig:fey} (b)),
no substantial difference is expected for \pagainst, compared to the
CS framework. Instead, 
a contribution due to the soft gluons emitted by the $c \bar{c}$ pair
forming the physical \(J/\psi\) state should be present. 
Hence, \pagainst is again sensitive to gluon fragmentation while \palong can 
shed light on the CO dynamics.
As NRQCD framework MC generators are not presently available for $e p$ 
collisions, only predictions of the CS model {\sc Herwig} MC are compared
to the data.

The momentum flow analysis was performed for different $p_T$ ranges.
All track quantities described in the following were measured in the 
laboratory frame at the reconstruction level.
Only primary vertex tracks with $p_T > 150$ \(\mev\) and $| \eta | < 1.75$ 
were selected.
The \(J/\psi\) decay muon tracks were discarded.
For each track whose component of momentum along the
\(J/\psi\) direction in the laboratory frame was positive, the component 
was included in \palong. If it was negative, it was included, in
absolute value, in \pagainst.
The data were restricted to $z > 0.3$ where the signal to background
ratio is highest.
The $W$ and $p_T$ ranges were $60 < W < 240$ \(\gev\) and $1 < p_T < 10$
\(\gev\), respectively.
The residual non-resonant background was subtracted for both
\pagainst and \palong variables using the shapes measured in the 
\(J/\psi\) side bands region and the normalisation obtained from the
signal extraction procedure described in Section~\ref{sec:cross-sections}.

The \pagainst (\palong) distribution, normalized to one, is shown
in Fig.~\ref{fig:pagainst} (\ref{fig:palong}). The prediction obtained
from the {\sc Herwig} MC simulation (including detector simulation) is also 
shown.
The \pagainst distribution of the MC simulation shows a softer drop 
from the first to the second momentum bin than that of the data. 
This situation is 
reversed for the higher momenta values where {\sc Herwig} predicts a steeper 
decrease than that observed in the data.
This behavior is seen for all $p_T$ regions.

For the \palong distribution, shown in Fig.~\ref{fig:palong}, a better
agreement is found between the {\sc Herwig} MC prediction and the data.

\section{Conclusions}

A measurement of the inelastic photoproduction of \(J/\psi\) and 
\(\psi^{\prime}\) mesons at HERA was presented.
The \(\psi^{\prime}\) to \(J/\psi\) cross section ratio was measured 
as a function of several kinematical observables.
The constant value of 0.25 predicted by the LO CS model is in reasonable 
agreement with the data.

Double differential cross sections of inelastic \(J/\psi\) 
photoproduction were measured.
A LO \(k_T\) calculation~\cite{epj:c71:1631} using CS terms alone gives,
within large normalisation uncertainties, 
a good description of the differential cross sections. However, for a 
better comparison with the data, a reduction of the theoretical uncertainties 
is very important.

A recent NLO calculation~\cite{prl:104:072001,*pr:d84:051501}, using CS 
and CO terms in the collinear approximation, gives a rough description of the 
double differential cross sections. The same calculation with only CS 
terms is in strong disagreement with the data. This leads to 
the conclusion that CO terms are an essential ingredient for this 
particular model.


Predictions of the {\sc Herwig} MC, which includes only CS processes,
were compared to the measured momentum flow along and against the \(J/\psi\) 
direction. {\sc Herwig} reproduces
the fall off of the momentum distribution against the \(J/\psi\) direction
as the momentum increases but fails to describe the exact shape of this 
distribution. A better description is obtained 
along the \(J/\psi\) direction.

\section*{Acknowledgments}
\vspace{0.3cm}
We appreciate the contributions to the construction and maintenance 
of the ZEUS detector of many people who are not listed as authors.  
The HERA machine group and the DESY computing staff are especially 
acknowledged for their success in providing excellent operation of 
the collider and the data analysis environment.  
We thank the DESY directorate for their strong support and encouragement.
It is a pleasure to thank S.~Baranov, M.~Butensch\"{o}n,
B.~Kniehl, A.~Lipatov, F.~Maltoni and N.~Zotov for helpful discussions 
and for providing their predictions.



%% file: ijphp-ref.tex
{
\ifzeusbst
  \bibliographystyle{./BiBTeX/bst/l4z_default}
\fi
\ifzdrftbst
  \bibliographystyle{./BiBTeX/bst/l4z_draft}
\fi
\ifzbstepj
  \bibliographystyle{./BiBTeX/user/l4z_epj-ICB}
\fi
\ifzbstnp
  \bibliographystyle{./BiBTeX/bst/l4z_np}
\fi
\ifzbstpl
  \bibliographystyle{./BiBTeX/bst/l4z_pl}
\fi
{\raggedright
\bibliography{./BiBTeX/user/syn.bib,%
              ./BiBTeX/user/myref.bib,%
              ./BiBTeX/bib/l4z_zeus.bib,%
              ./BiBTeX/bib/l4z_h1.bib,%
              ./BiBTeX/bib/l4z_articles.bib,%
              ./BiBTeX/bib/l4z_books.bib,%
              ./BiBTeX/bib/l4z_conferences.bib,%
              ./BiBTeX/bib/l4z_misc.bib,%
              ./BiBTeX/bib/l4z_preprints.bib}}
}
\vfill\eject

%% file: ijphp-tab.tex
%
%
%
%
\newpage
\begin{table}
\begin{footnotesize}
\begin{center}
\renewcommand{\arraystretch}{1.3}\renewcommand{\arraystretch}{1.3}
\begin{tabular}[t]{|c|c|c|} \hline
\multicolumn{3}{|c|}{$\psi^{'}$ to $J/\psi$ cross section ratio: kinematic range} \\

\hline \hline
60 $< W <$ 190 GeV  &  $p_{T} > 0$ GeV  & 0.55 $< z <$ 0.9 \\
\hline \hline 

\multicolumn{3}{|c|}{Differential cross sections: kinematic range } \\
\hline \hline
60 $< W <$ 240 GeV  &  $p_{T} > 1$ GeV  & 0.1 $< z <$ 0.9 \\
\hline \hline

\multicolumn{3}{|c|}{Momentum flow: kinematic range} \\
\hline \hline
60 $< W <$ 240 GeV  &  $1 < p_{T} < 10$ GeV  & 0.3 $< z <$ 0.9 \\
\hline 
\end{tabular} 
\end{center}
\caption{The different kinematic regions used in the measurement of the
 $\psi^{'}$ to $J/\psi$ cross section ratio,
 $J/\psi$ differential cross sections
and momentum flow along and against the $J/\psi$ direction.}
\label{tab:kinrange}
\end{footnotesize}
\end{table}

\newpage
\begin{table}
\begin{footnotesize}
\begin{center}
\renewcommand{\arraystretch}{1.3}\renewcommand{\arraystretch}{1.3}
\begin{tabular}[t]{|c|c|c|} \hline
\multicolumn{1}{|c|}{$p_T$ range} &
\multicolumn{1}{c|}{$\langle p_T \rangle$} & 
\multicolumn{1}{c|}{$\sigma(\psi^{'})/\sigma(J/\psi)$} \\
\multicolumn{1}{|c|}{(GeV)} & \multicolumn{1}{c|}{(GeV)} & \multicolumn{1}{c|}{} \\
\hline \hline
\(0.0 - 1.0\)  &  0.63 & $ 0.262 \pm 0.043 ^{+0.003}_{-0.014}$ \\
\(1.0 - 1.75\) &  1.35 & $ 0.317 \pm 0.049 ^{+0.010}_{-0.005}$ \\
\(1.75 - 5.0\) &  2.68 & $ 0.263 \pm 0.041 ^{+0.030}_{-0.002}$ \\ 
\hline \hline 

\multicolumn{1}{|c|}{$W$ range} &
\multicolumn{1}{c|}{$\langle W \rangle$} & 
\multicolumn{1}{c|}{$\sigma(\psi^{'})/\sigma(J/\psi)$} \\
\multicolumn{1}{|c|}{(GeV)} & \multicolumn{1}{c|}{(GeV)} & \multicolumn{1}{c|}{} \\
\hline \hline
\(60 - 95\)    &  81.42 & $ 0.368 \pm 0.054 ^{+0.052}_{-0.042}$ \\
\(95 - 120\)   &  108.03& $ 0.409 \pm 0.057 ^{+0.006}_{-0.015}$ \\
\(120 - 190\)  &  149.11& $ 0.218 \pm 0.040 ^{+0.026}_{-0.015}$ \\
\hline \hline 

\multicolumn{1}{|c|}{$z$ range} &
\multicolumn{1}{c|}{$\langle z \rangle$} & 
\multicolumn{1}{c|}{$\sigma(\psi^{'})/\sigma(J/\psi)$} \\
\multicolumn{1}{|c|}{} & \multicolumn{1}{c|}{} & \multicolumn{1}{c|}{} \\
\hline \hline

\(0.55 - 0.70\) &  0.62 & $ 0.250 \pm 0.043 ^{+0.014}_{-0.015}$ \\
\(0.70 - 0.80\)  &  0.75 & $ 0.289 \pm 0.040 ^{+0.007}_{-0.019}$ \\
\(0.80 - 0.90\)  &  0.85 & $ 0.344 \pm 0.054 ^{+0.036}_{-0.008}$ \\
\hline 
\end{tabular} 
\end{center}
\caption{Cross section ratio of $\psi^{'}$ to $J/\psi$ as a function of
$p_T$, $W$ and $z$ in the kinematic 
region $60  < W <  190$ GeV and $0.55 < z < 0.9$.
In the quoted ratios, the first uncertainty is statistical and the
second is systematic.
}
\label{tab:ratio}
\end{footnotesize}
\end{table}

\newpage
\begin{table}
\begin{scriptsize}
\begin{center} 
\renewcommand{\arraystretch}{1.3}
\begin{tabular}[t]{|c|c|c|c|c|} \hline
\multicolumn{1}{|c|}{$z$ range} &
\multicolumn{1}{c|}{$p^2_T$ range} & \multicolumn{1}{c|}{$\langle p^2_T \rangle$} &
\multicolumn{1}{c|}{$d\sigma/dp^2_T $} & 
\multicolumn{1}{c|}{$d\sigma(b \rightarrow J/\psi)/dp^2_T$} \\
\multicolumn{1}{|c|}{} & \multicolumn{1}{c|}{(GeV$^2$)} & \multicolumn{1}{c|}{(GeV$^2$)} &
\multicolumn{1}{c|}{(nb/GeV$^2$)}   & \multicolumn{1}{c|}{(nb/GeV$^2$)}   \\
\hline \hline
\(0.10 -0.30\)&\(1.0 - 2.0 \)  & 1.46 & $ 1.03    \pm 0.13    ^{+0.18}_{-0.15}   $  & 0.05 \\
&\(2.0 - 3.0 \)  & 2.47 & $ 0.86    \pm 0.12    ^{+0.10}_{-0.17}   $ & 0.04\\
&\(3.0 - 4.5 \)  & 3.67 & $ 0.410   \pm 0.079   ^{+0.055}_{-0.068}  $ & 0.029\\
&\(4.5 - 7.0 \)  & 5.64 & $ 0.127   \pm 0.047   ^{+0.020}_{-0.027}  $ & 0.018\\
&\(7.0 - 10.0 \) & 8.37 & $ 0.052   \pm 0.030   ^{+0.022}_{-0.008}  $ & 0.012\\
&\(10.0 - 14.0 \)& 11.62& $ 0.056   \pm 0.017   ^{+0.009}_{-0.006}  $ & 0.007\\
&\(14.0 - 20.0 \)& 16.34& $ 0.0329  \pm 0.0081  ^{+0.0029}_{-0.0066}$ & 0.0035\\
&\(20.0 - 40.0 \)& 26.50& $ 0.0069  \pm 0.0018  ^{+0.0010}_{-0.0008}$ & 0.0012\\
&\(40.0- 100.0 \)& 56.69& $ 0.00092 \pm 0.00037 ^{+0.00018}_{-0.00026}$ & 0.00013\\  
\hline 
\(0.30 - 0.45 \)&\(1.0 - 2.0 \)  & 1.47 & $ 1.32      \pm 0.10     ^{+0.21}_{-0.16}  $  & 0.02\\
&\(2.0 - 3.0 \)  & 2.45 & $ 0.823     \pm 0.081    ^{+0.094}_{-0.101} $ & 0.018\\
&\(3.0 - 4.5 \)  & 3.70 & $ 0.492     \pm 0.060    ^{+0.071}_{-0.075} $ & 0.013\\
&\(4.5 - 7.0 \)  & 5.64 & $ 0.190     \pm 0.032    ^{+0.024}_{-0.028} $  & 0.010\\
&\(7.0 - 10.0 \) & 8.35 & $ 0.111     \pm 0.019    ^{+0.014}_{-0.013} $  & 0.006\\
&\(10.0 - 14.0 \)& 11.77& $ 0.062     \pm 0.011    ^{+0.010}_{-0.007} $ & 0.004\\
&\(14.0 - 20.0 \)& 16.49& $ 0.0349    \pm 0.0052   ^{+0.0030}_{-0.0035}$ & 0.0021\\
&\(20.0 - 40.0 \)& 27.96& $ 0.0065    \pm 0.0012   ^{+0.0009}_{-0.0008}$ & 0.0007\\
&\(40.0- 100.0 \)& 54.05& $ 0.00095   \pm 0.00019  ^{+0.00014}_{-0.00007}$ & 0.00009\\  
\hline
\(0.45 - 0.60 \)&\(1.0 - 2.0 \)  & 1.45 & $ 2.20       \pm 0.09      ^{+0.25}_{-0.25}   $ & - \\
&\(2.0 - 3.0 \)  & 2.47 & $ 1.38       \pm 0.08      ^{+0.15}_{-0.16}   $ & - \\
&\(3.0 - 4.5 \)  & 3.69 & $ 0.84       \pm 0.05      ^{+0.12}_{-0.10}   $ & - \\
&\(4.5 - 7.0 \)  & 5.65 & $ 0.424      \pm 0.029     ^{+0.054}_{-0.058}  $ & - \\
&\(7.0 - 10.0 \) & 8.35 & $ 0.249      \pm 0.017     ^{+0.029}_{-0.029}  $ & - \\
&\(10.0 - 14.0 \)& 11.79& $ 0.121      \pm 0.010     ^{+0.013}_{-0.013}  $ & - \\
&\(14.0 - 20.0 \)& 16.60& $ 0.0505     \pm 0.0048    ^{+0.0046}_{-0.0051}$ & 0.0007\\
&\(20.0 - 40.0 \)& 26.70& $ 0.0106     \pm 0.0011    ^{+0.0009}_{-0.0009}$ & 0.0004\\
&\(40.0- 100.0 \)& 55.86& $ 0.00122    \pm 0.00020   ^{+0.00013}_{-0.00015}$ & 0.00003\\  
\hline
\( 0.60 - 0.75 \)&\(1.0 - 2.0 \)  & 1.45 & $ 2.80    \pm 0.10     ^{+0.35}_{-0.32}   $ & - \\
&\(2.0 - 3.0 \)  & 2.47 & $ 2.07    \pm 0.09     ^{+0.23}_{-0.23}   $ & - \\
&\(3.0 - 4.5 \)  & 3.70 & $ 1.10    \pm 0.05     ^{+0.13}_{-0.13}   $ & - \\
&\(4.5 - 7.0 \)  & 5.60 & $ 0.680   \pm 0.030    ^{+0.084}_{-0.084} $ & - \\
&\(7.0 - 10.0 \) & 8.38 & $ 0.286   \pm 0.017    ^{+0.031}_{-0.036} $ & - \\
&\(10.0 - 14.0 \)& 11.93& $ 0.153   \pm 0.010    ^{+0.015}_{-0.016}$ & - \\
&\(14.0 - 20.0 \)& 16.92& $ 0.0532  \pm 0.0044   ^{+0.0051}_{-0.0050}$ & - \\
&\(20.0 - 40.0 \)& 27.0 \hspace{0.7 mm} & $ 0.0123  \pm 0.0011   ^{+0.0011}_{-0.0011}$ & - \\
&\(40.0- 100.0 \)& 55.77& $ 0.00112 \pm 0.00018  ^{+0.00010}_{-0.00022}$ & - \\  
\hline
\(0.75 - 0.90 \) &\(1.0 - 2.0 \)  & 1.45 & $ 2.39     \pm 0.13      ^{+0.51}_{-0.35}   $ & - \\
&\(2.0 - 3.0 \)  & 2.45 & $ 1.77     \pm 0.11      ^{+0.44}_{-0.23}   $ & - \\
&\(3.0 - 4.5 \)  & 3.66 & $ 1.17     \pm 0.07      ^{+0.16}_{-0.15}   $ & - \\
&\(4.5 - 7.0 \)  & 5.64 & $ 0.716    \pm 0.039     ^{+0.087}_{-0.092}  $ & - \\
&\(7.0 - 10.0 \) & 8.31 & $ 0.369    \pm 0.023     ^{+0.042}_{-0.045}  $ & - \\
&\(10.0 - 14.0 \)& 11.77& $ 0.166    \pm 0.012     ^{+0.016}_{-0.020}  $ & - \\
&\(14.0 - 20.0 \)& 16.66& $ 0.0650   \pm 0.0058    ^{+0.0053}_{-0.0087}$ & - \\
&\(20.0 - 40.0 \)& 26.22& $ 0.0139   \pm 0.0013    ^{+0.0012}_{-0.0025}$ & - \\
&\(40.0- 100.0 \)& 54.0 \hspace{0.8 mm}& $ 0.00093  \pm 0.00018   ^{+0.00007}_{-0.00018}$ & - \\  
\hline
\end{tabular} 
\end{center}
\caption{Measured $J/\psi$ differential photoproduction cross sections in the 
kinematic region $0.1  < z <  0.9$ and $60  < W <  240$ GeV as a function 
of the squared transverse momentum of the  $J/\psi$ mesons in bins of
inelasticity $z$. 
In the quoted cross sections, the first uncertainty is statistical and the second
is systematic.
The bin center values $\langle p^2_T\rangle $ 
and the expected, but not subtracted, beauty contribution (estimated through 
the {\sc Pythia} MC)
are also given in the table. The beauty contribution is only given
when its value is above 1\% with respect to the corresponding 
measured differential photoproduction cross section.}
\label{tab:2diffpt}
\end{scriptsize}
\end{table}

\newpage
\begin{table}
\begin{footnotesize}
\begin{center}
\renewcommand{\arraystretch}{1.35}
\begin{tabular}[t]{|c|c|c|c|c|} \hline
\multicolumn{1}{|c|}{$p_T$ range} &
\multicolumn{1}{c|}{$z$ range} & \multicolumn{1}{c|}{$\langle z \rangle$} &  
\multicolumn{1}{c|}{$d\sigma/dz$} &
\multicolumn{1}{c|}{$d\sigma(b \rightarrow J/\psi)/dz$} \\
\multicolumn{1}{|c|}{(GeV)} & \multicolumn{1}{c|}{} & \multicolumn{1}{c|}{} &
\multicolumn{1}{c|}{(nb)}   & \multicolumn{1}{c|}{(nb)}   \\
\hline \hline
\(1.0 - 2.0 \) &\(0.10 - 0.30 \) & 0.21 & $11.5 \pm 1.0^{+1.5}_{-1.9}$  & 0.6 \\
& \(0.30 - 0.45 \) & 0.37 & $17.3 \pm 1.0^{+2.3}_{-2.2}$ & 0.4 \\
& \(0.45 - 0.60 \) & 0.52 & $29.9 \pm 0.9^{+3.3}_{-3.4}$ & - \\
& \(0.60 - 0.75 \) & 0.67 & $40.2 \pm 1.0^{+4.4}_{-4.5}$ & - \\
& \(0.75 - 0.90 \) & 0.82 & $36.6 \pm 1.2^{+6.6}_{-4.9}$ & - \\
\hline 
\(2.0 - 3.0 \) &\(0.10 - 0.30 \) & 0.21 & $1.94 \pm 0.78^{+0.31}_{-0.45}$ & 0.43 \\
&\(0.30 - 0.45 \) & 0.37 & $6.42 \pm 0.71^{+0.78}_{-0.81}$  & 0.28 \\
&\(0.45 - 0.60 \) & 0.52 & $12.4 \pm 0.6 ^{+1.5}_{-1.5}$ &  0.8 \\
&\(0.60 - 0.75 \) & 0.67 & $18.6 \pm 0.6 ^{+2.2}_{-2.2}$ & - \\
&\(0.75 - 0.90 \) & 0.82 & $19.4 \pm 0.8 ^{+2.4}_{-2.3}$ & - \\
\hline 
\(3.0 - 4.5 \) &\(0.10 - 0.30 \) & 0.20 & $2.55 \pm 0.47^{+0.36}_{-0.28}$ & 0.30 \\
&\(0.30 - 0.45 \) & 0.38 & $3.51 \pm 0.41^{+0.37}_{-0.37}$ & 0.26 \\
&\(0.45 - 0.60 \) & 0.52 & $6.61 \pm 0.37^{+0.66}_{-0.72}$ & 0.08 \\
&\(0.60 - 0.75 \) & 0.68 & $7.79 \pm 0.35^{+0.74}_{-0.75}$ & - \\
&\(0.75 - 0.90 \) & 0.82 & $9.29 \pm 0.46^{+0.84}_{-1.15}$ & - \\
\hline 
$ > 4.5$ &\(0.10 - 0.30 \) & 0.21 & $1.01 \pm 0.20^{+0.10}_{-0.13}$ & 0.16 \\
&\(0.30 - 0.45 \) & 0.38 & $1.31 \pm 0.18^{+0.15}_{-0.12}$ & 0.24 \\
&\(0.45 - 0.60 \) & 0.52 & $1.98 \pm 0.17^{+0.16}_{-0.16}$ & 0.08 \\
&\(0.60 - 0.75 \) & 0.67 & $2.11 \pm 0.16^{+0.18}_{-0.19}$ & - \\
&\(0.75 - 0.90 \) & 0.82 & $2.16 \pm 0.18^{+0.18}_{-0.49}$ & - \\
\hline 
\end{tabular} 
\end{center}
\caption{Measured $J/\psi$ differential photoproduction cross sections in the 
kinematic region $p_T > 1$ GeV and $60  < W <  240$ GeV as a function 
of the inelasticity $z$ in bins of transverse momentum of the $J/\psi$ meson. 
In the quoted cross sections, the first uncertainty is statistical and the second
is systematic.
The bin center values $\langle z \rangle$ and the expected, but not subtracted, beauty 
contribution (estimated through the {\sc Pythia} MC) are also given in the 
table.
For further details see Table \ref{tab:2diffpt}.}
\label{tab:2diffz}
\end{footnotesize}
\end{table}

%% file: ijphp-fig.tex
\begin{figure}
\unitlength1cm  \begin{picture}(15.,7.5)
\includegraphics{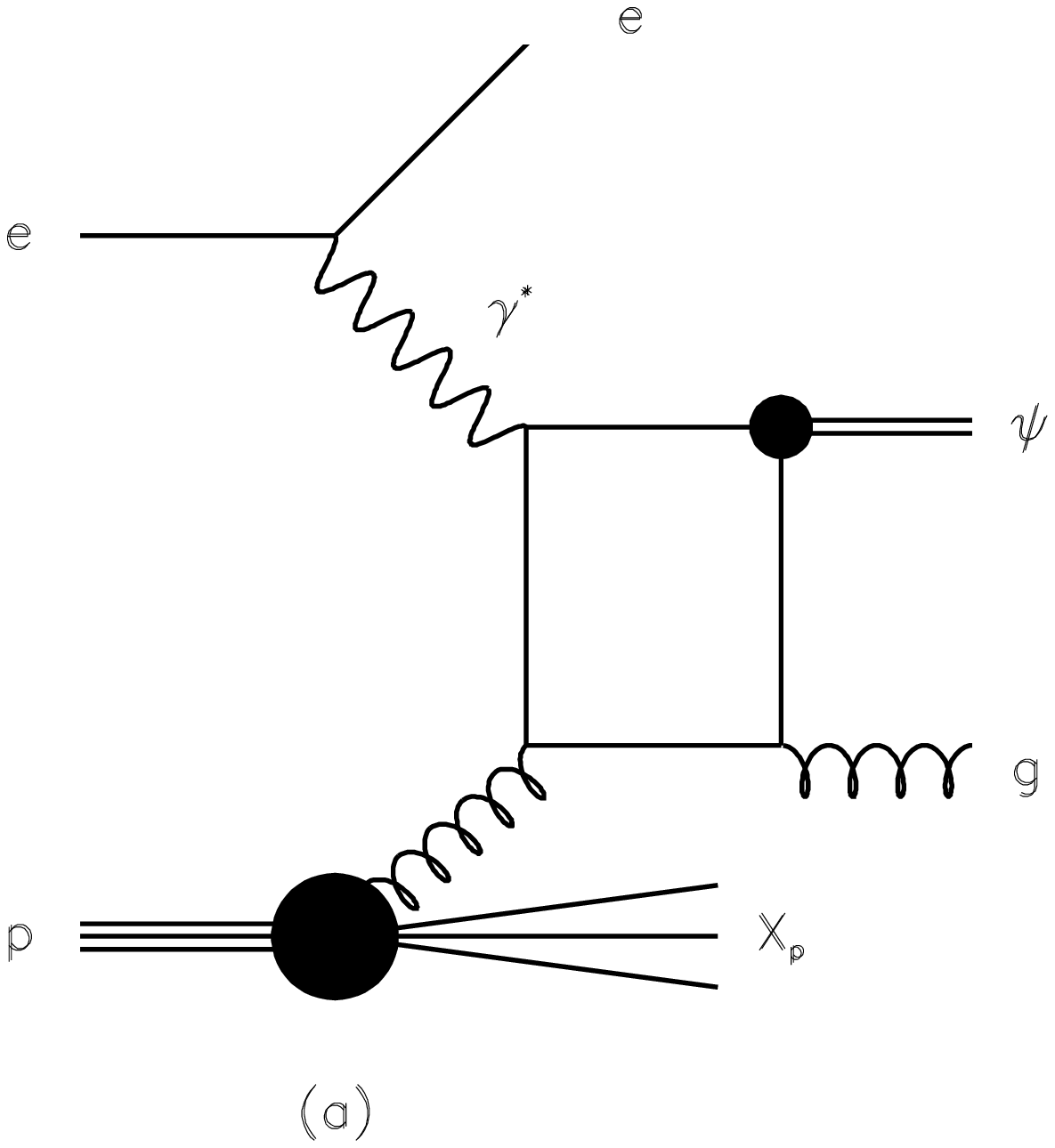}
\includegraphics{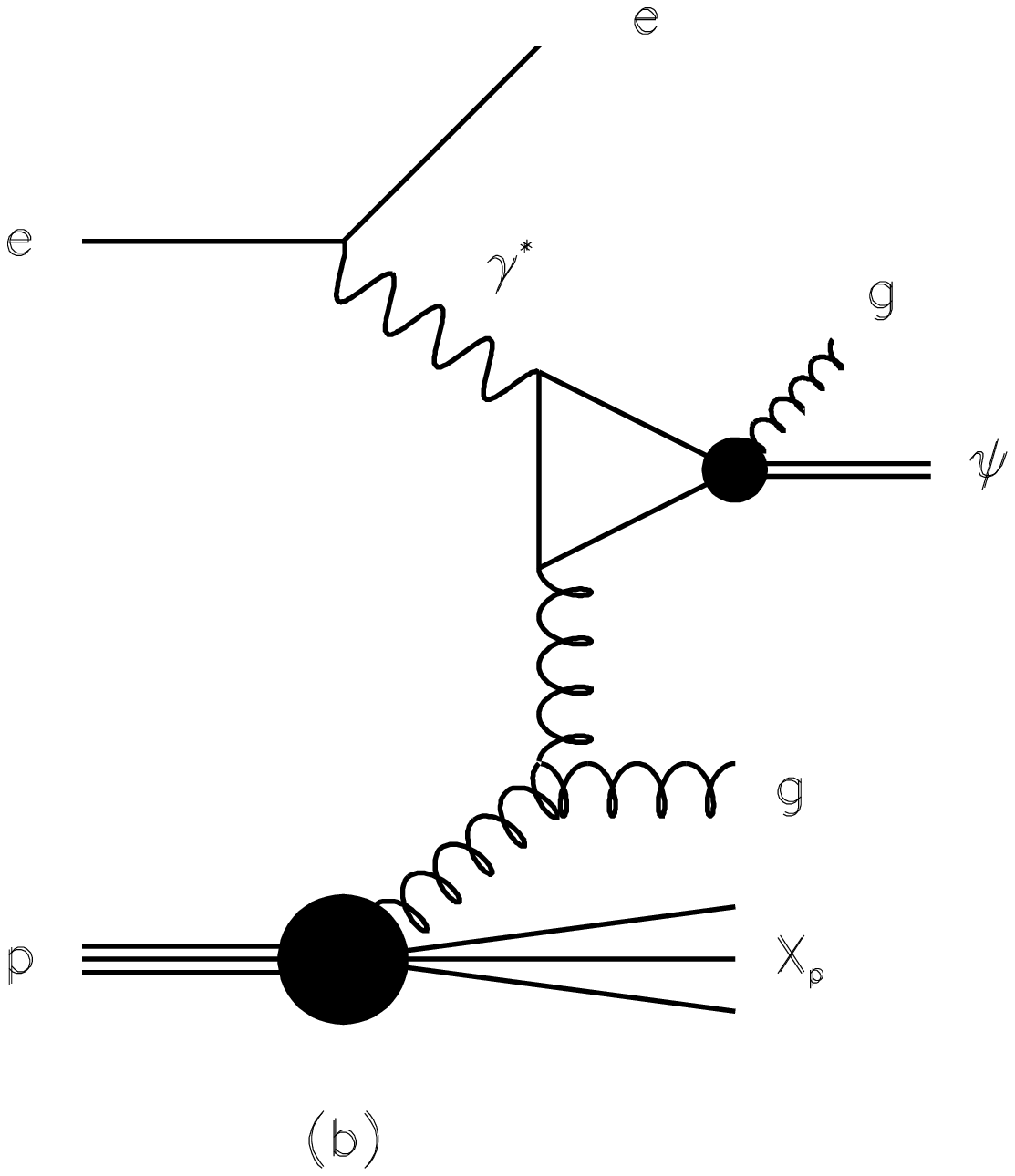}
\end{picture}
\caption{
Examples of direct photon-processes at leading-order in (a) the 
colour-singlet and (b) the colour-octet frameworks.}
\label{fig:fey}
\end{figure}

\begin{figure}
\unitlength1cm  \begin{picture}(15.,8.5)
\includegraphics{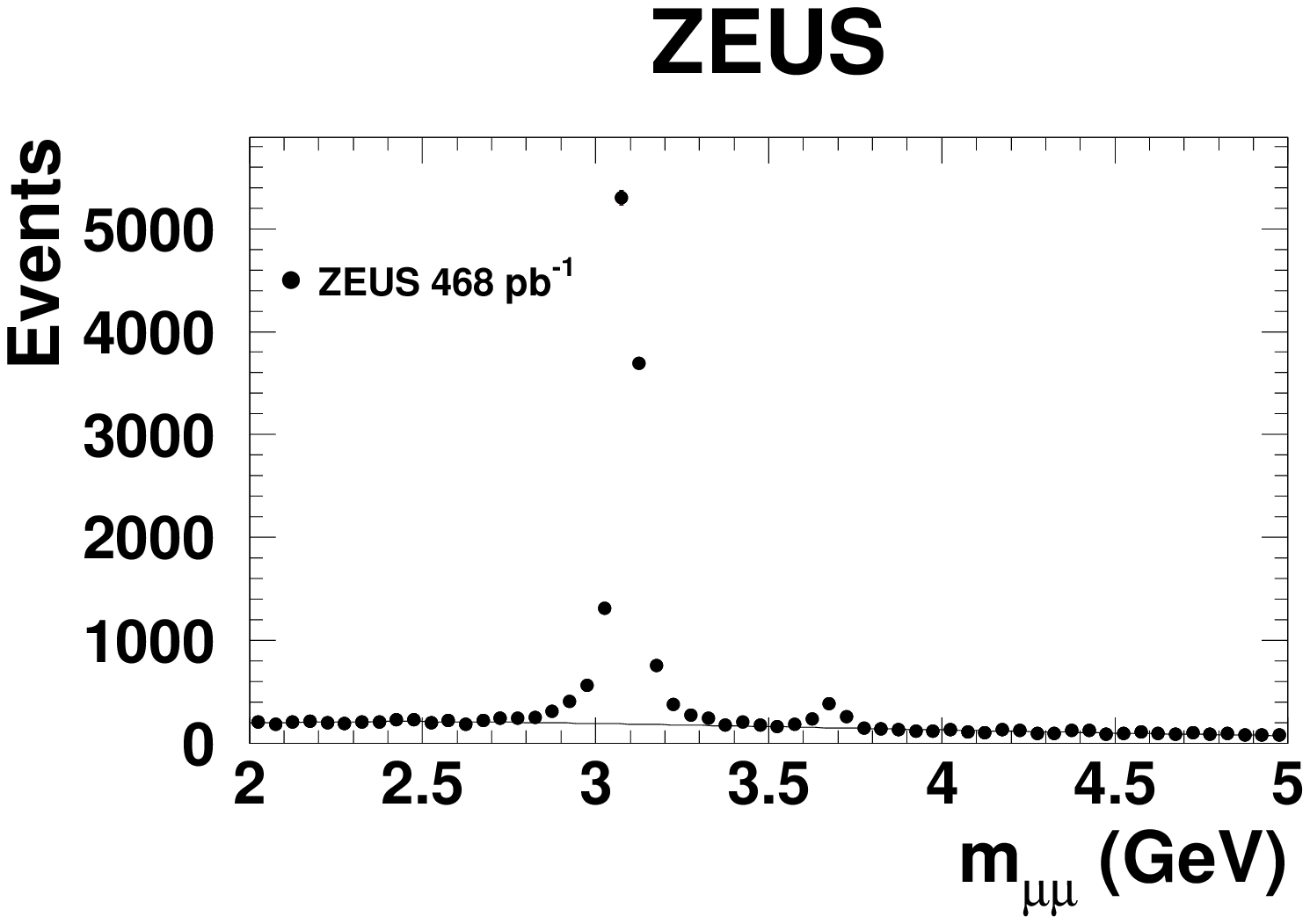}
\includegraphics{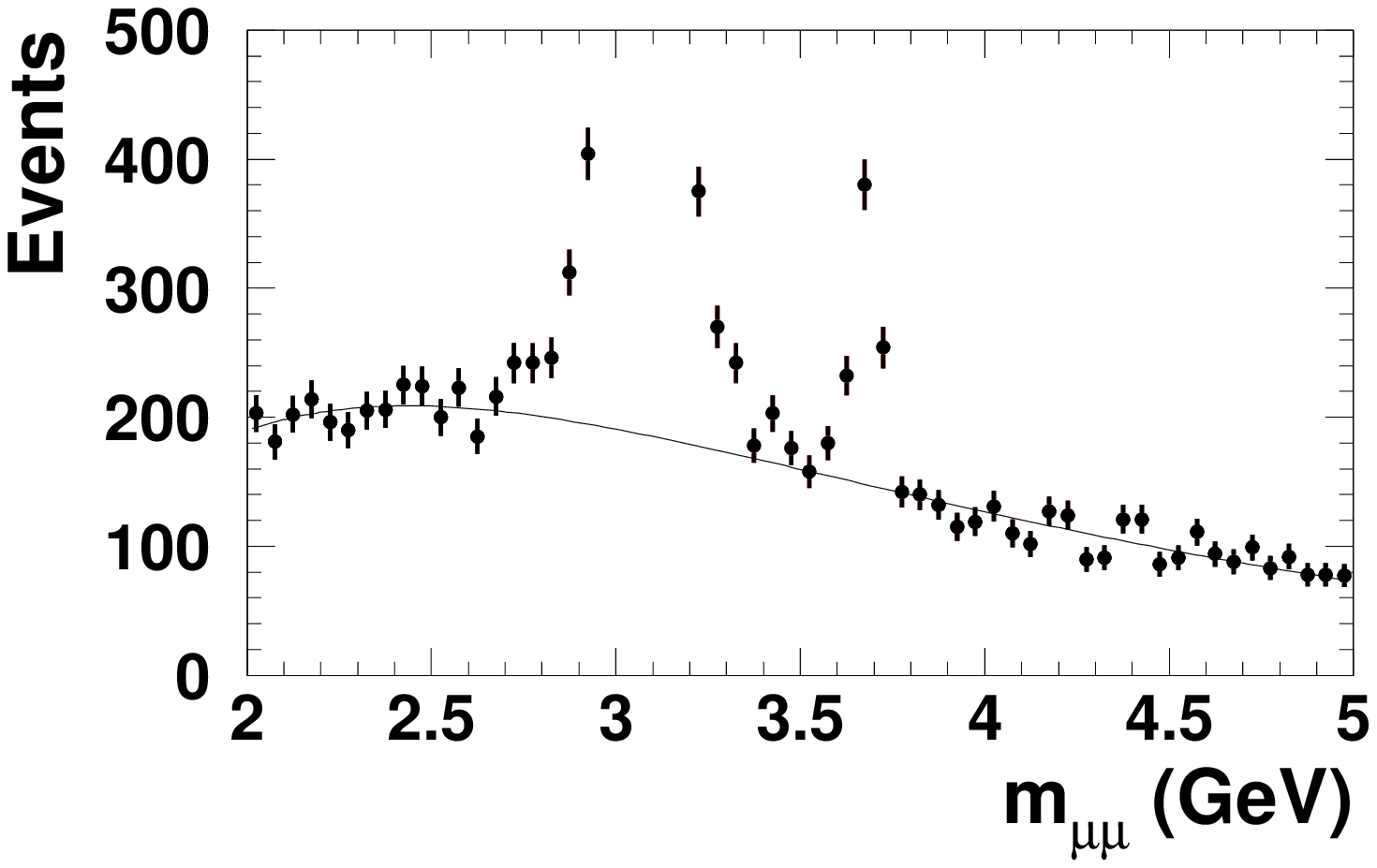}
\end{picture}
\caption{Invariant-mass distribution, $m_{\mu \mu}$, in the kinematic
region $0.55 < z < 0.9$ and $60 < W < 190$ GeV.
The continuous line shows the estimated background contribution 
(for further details see the text).
The right insert highlights the \(\psi^{\prime}\) mass peak.}
\label{fig:mmumu-pap}
\end{figure}

\begin{figure}[hbpt!]
\begin{center}
\unitlength1cm
\includegraphics[width=1.0\textwidth]{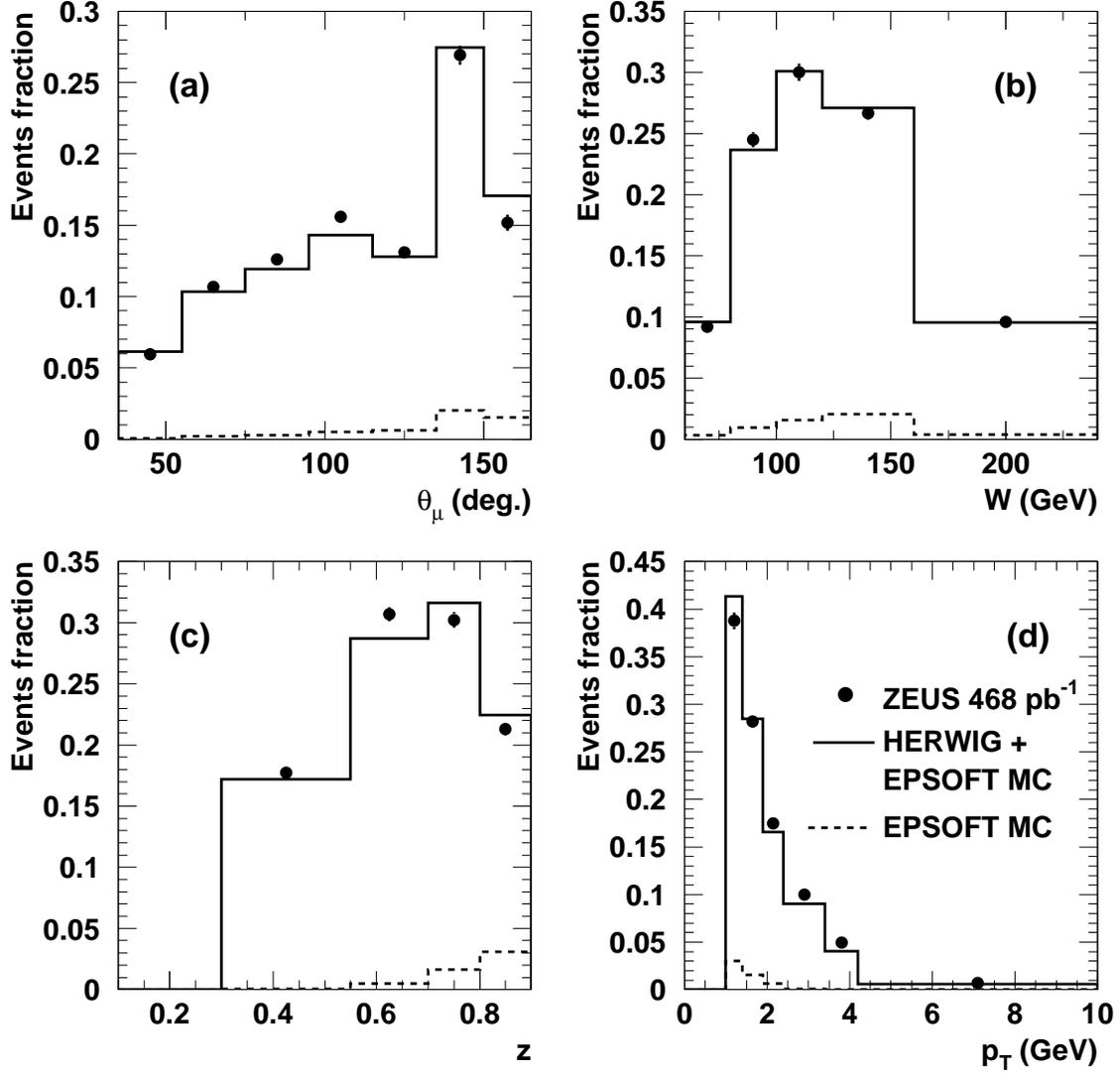}
\caption{\(J/\psi\) events fraction measured in the kinematic
region $0.3 < z < 0.9$, $60 < W < 240$ GeV and $p_T > 1$ GeV 
as a function of (a) the polar angle $\theta_{\mu}$ of the
muon tracks, (b) $W$, (c) the inelasticity $z$ and (d) the
\(J/\psi\) \(p_T\).
The data are shown as points. The error bars are the statistical
uncertainties.
The sum of the {\sc Herwig} and {\sc Epsoft} MC predictions, 
according to the relative fraction described in the text and normalised 
to the data are also shown (continuous lines).
The {\sc Epsoft} MC component is shown separately (dashed lines).}
\label{fig:ccpl-pap}
\end{center}
\end{figure}

\begin{figure}[hbpt!]
\begin{center}
\unitlength1cm
\includegraphics[width=1.0\textwidth]{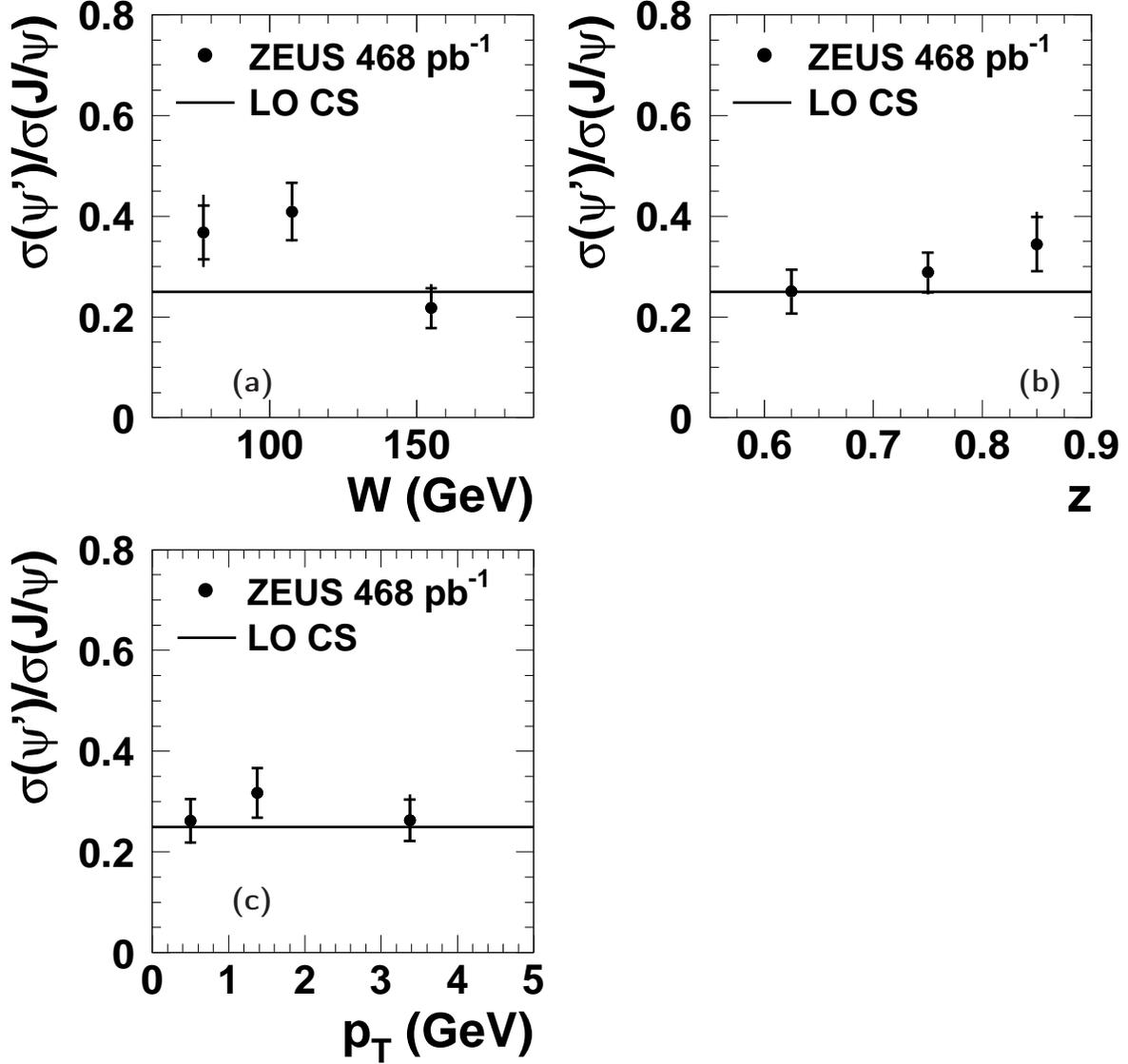}
\put(-13.,9.5){\textsf{\textbf{ (a) }}}
\put(-2.,9.5){\textsf{\textbf{ (b) }}}
\put(-13.,2.25){\textsf{\textbf{ (c) }}}
\caption{\(\psi^{\prime}\) to  \(J/\psi\) photoproduction cross section 
ratio measured in the kinematic region $0.55 < z < 0.9$ and 
$60 < W < 190$ GeV as a function of (a) $W$, (b) the inelasticity $z$
and (c) $p_T$.
The data are shown as points. The inner error bars are the statistical
uncertainties, while the outer error bars show the statistical and
systematic uncertainties added in quadrature.
The leading-order colour-singlet model expectation (horizontal lines) is 
also shown.}
\label{fig:2sto1s}
\end{center}
\end{figure}

\begin{figure}[hbpt!]
\unitlength1cm
\includegraphics[width=1.0\textwidth]{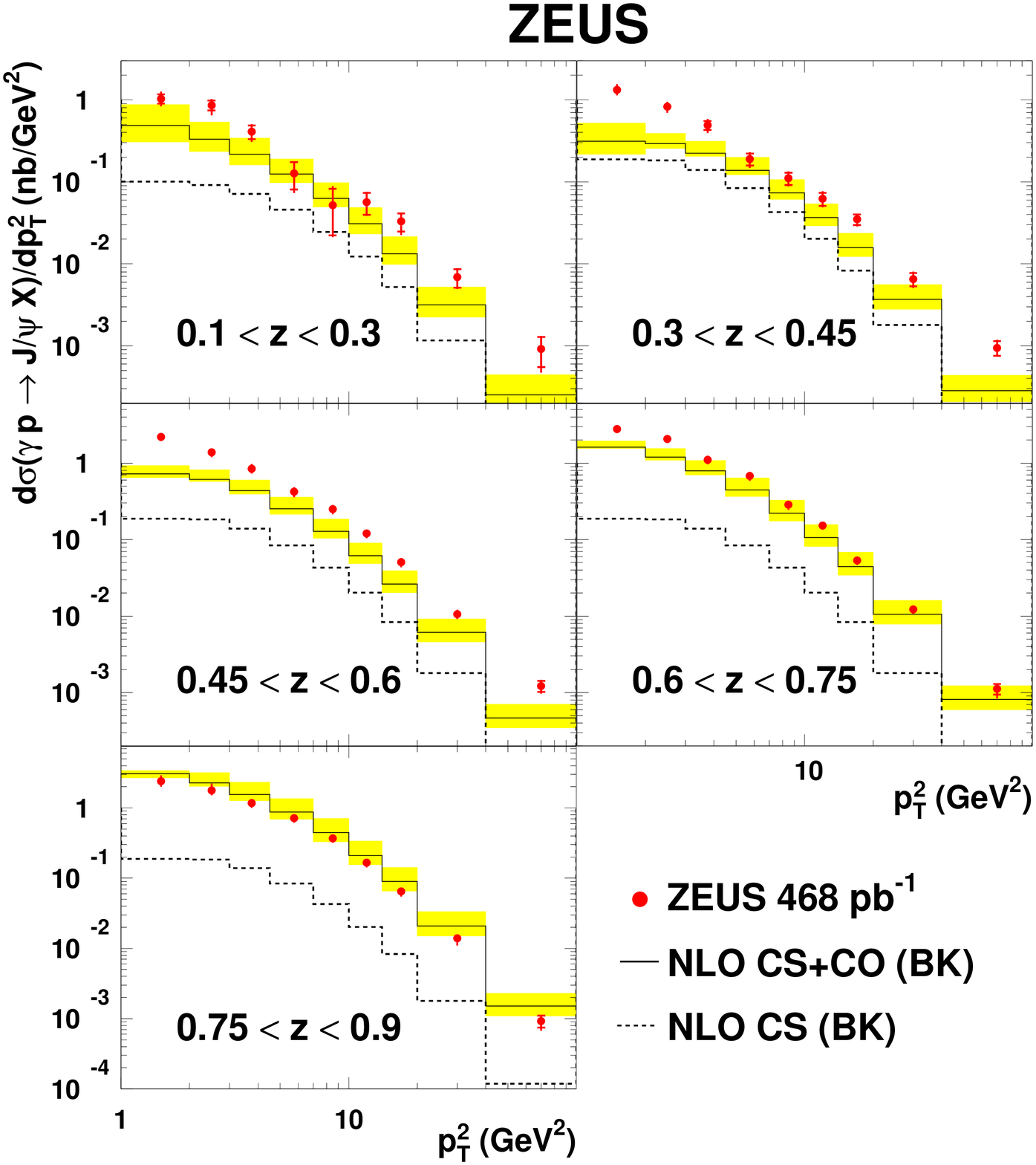}
\caption{Differential cross sections $d\sigma/dp_T^2$ measured in 5 
different $z$ ranges.
The measurement is performed in the kinematic region 
$60 < W < 240$ GeV and $p_T > 1$ GeV.
The data are shown as points.
The inner (outer) error bars represent the statistical (total) 
uncertainties.
The solid lines show the NLO CS+CO (BK) 
prediction~\protect\cite{prl:104:072001,*pr:d84:051501} obtained 
in the non-relativistic QCD framework. The uncertainties are indicated
by the band. The colour-singlet model contribution is presented separately
as the dashed lines.}
\label{fig:dsdpt2-nlocs+co}
\end{figure}

\begin{figure}[hbpt!]
\unitlength1cm
\includegraphics[width=1.0\textwidth]{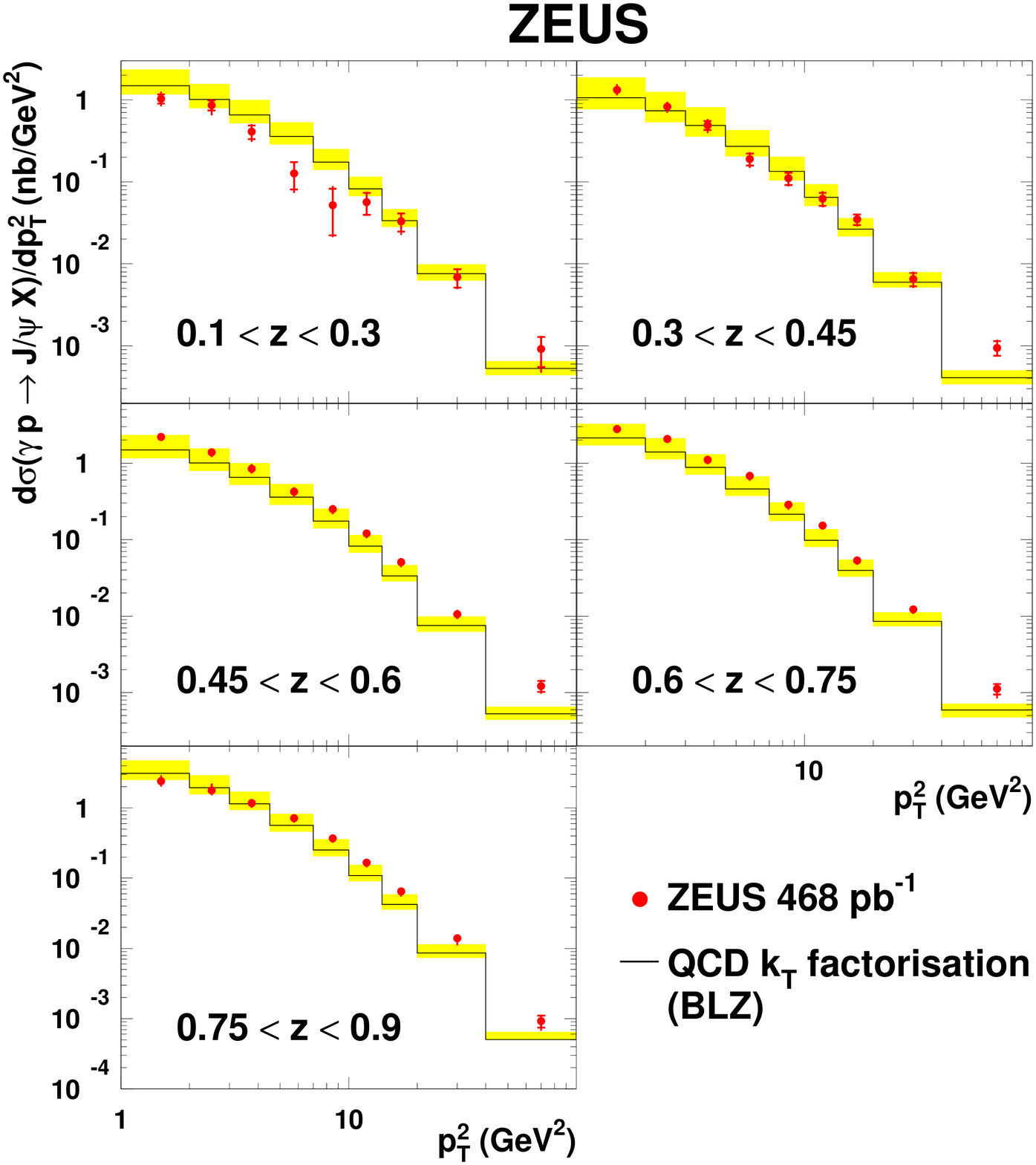}
\caption{Differential cross sections $d\sigma/dp_T^2$ measured in 5 
different $z$ ranges.
The measurement is performed in the kinematic region 
$60 < W < 240$ GeV and $p_T > 1$ GeV.
The data are shown as points.
The inner (outer) error bars represent the statistical (total) 
uncertainties.
The solid lines show the \(k_T\)--factorisation (BLZ)
prediction~\protect\cite{epj:c27:87,epj:c71:1631}.
The uncertainties are indicated by the band.}
\label{fig:dsdpt2-kt}
\end{figure}

\begin{figure}[hbpt!]
\unitlength1cm
\includegraphics[width=0.95\textwidth]{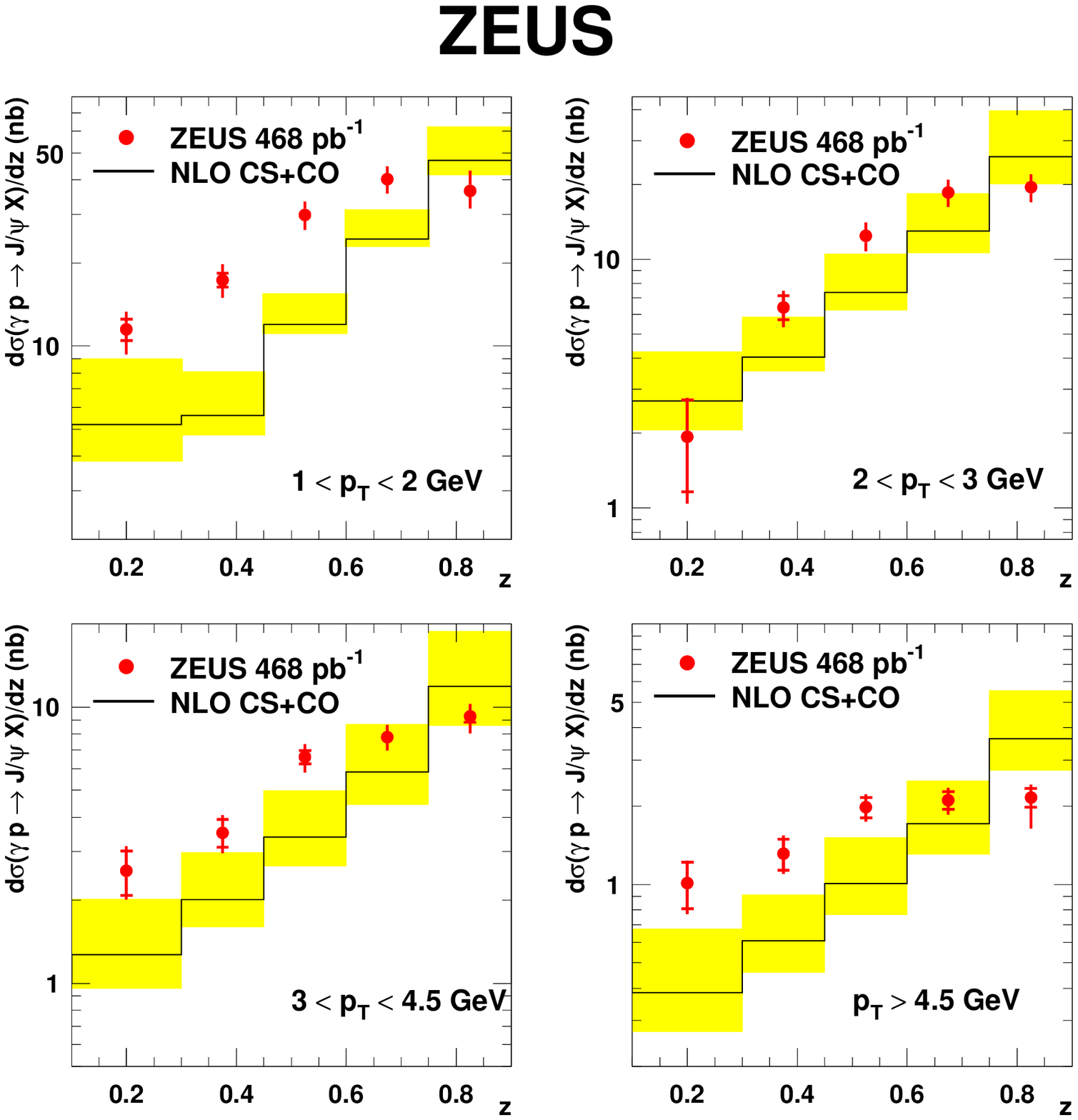}
\caption{Differential $J/\psi$ cross sections $d\sigma/dz$ measured in 4
different $p_T$ ranges.
The measurement is performed in the kinematic region 
$60 < W < 240$ GeV and $0.1 < z < 0.9$.
The data are shown as points.
The inner (outer) error bars represent the statistical (total) 
uncertainties.
The solid lines show the NLO CS+CO (BK) 
prediction~\protect\cite{prl:104:072001,*pr:d84:051501} obtained 
in the non-relativistic QCD framework. The uncertainties are indicated
by the band.}
\label{fig:dsdz-nlocs+co}
\end{figure}

\begin{figure}[hbpt!]
\unitlength1cm
\includegraphics[width=0.95\textwidth]{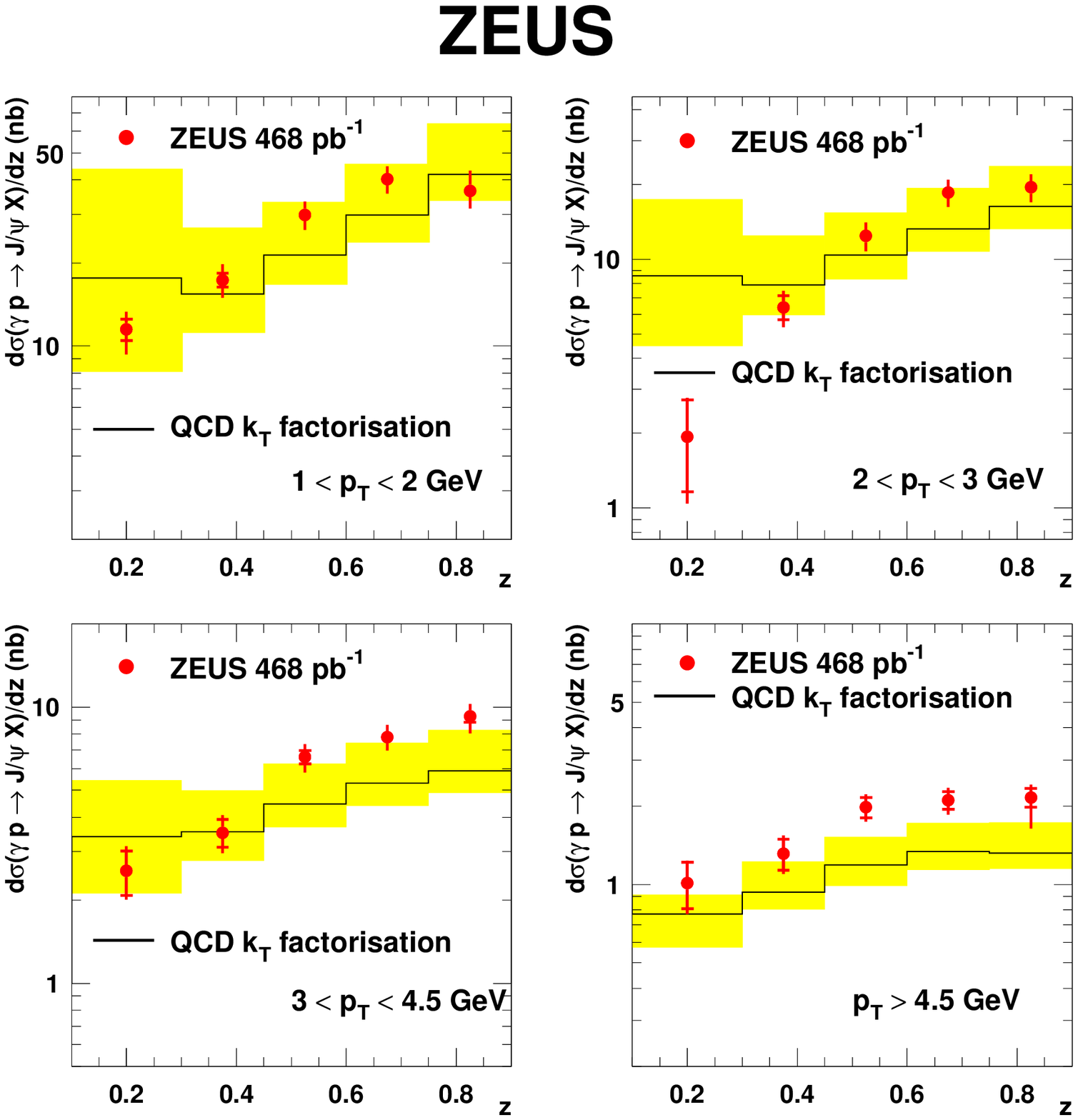}
\caption{Differential $J/\psi$ cross sections $d\sigma/dz$ measured in 4
different $p_T$ ranges.
The measurement is performed in the kinematic region 
$60 < W < 240$ GeV and $0.1 < z < 0.9$.
The data are shown as points.
The inner (outer) error bars represent the statistical (total) 
uncertainties.
The solid lines show the \(k_T\)--factorisation (BLZ)
prediction~\protect\cite{epj:c27:87,epj:c71:1631}.
The uncertainties are indicated by the band.}
\label{fig:dsdz-kt}
\end{figure}

\begin{figure}[hbpt!]
\unitlength1cm
\hspace{-0.75cm}
\includegraphics[width=1.05\textwidth]{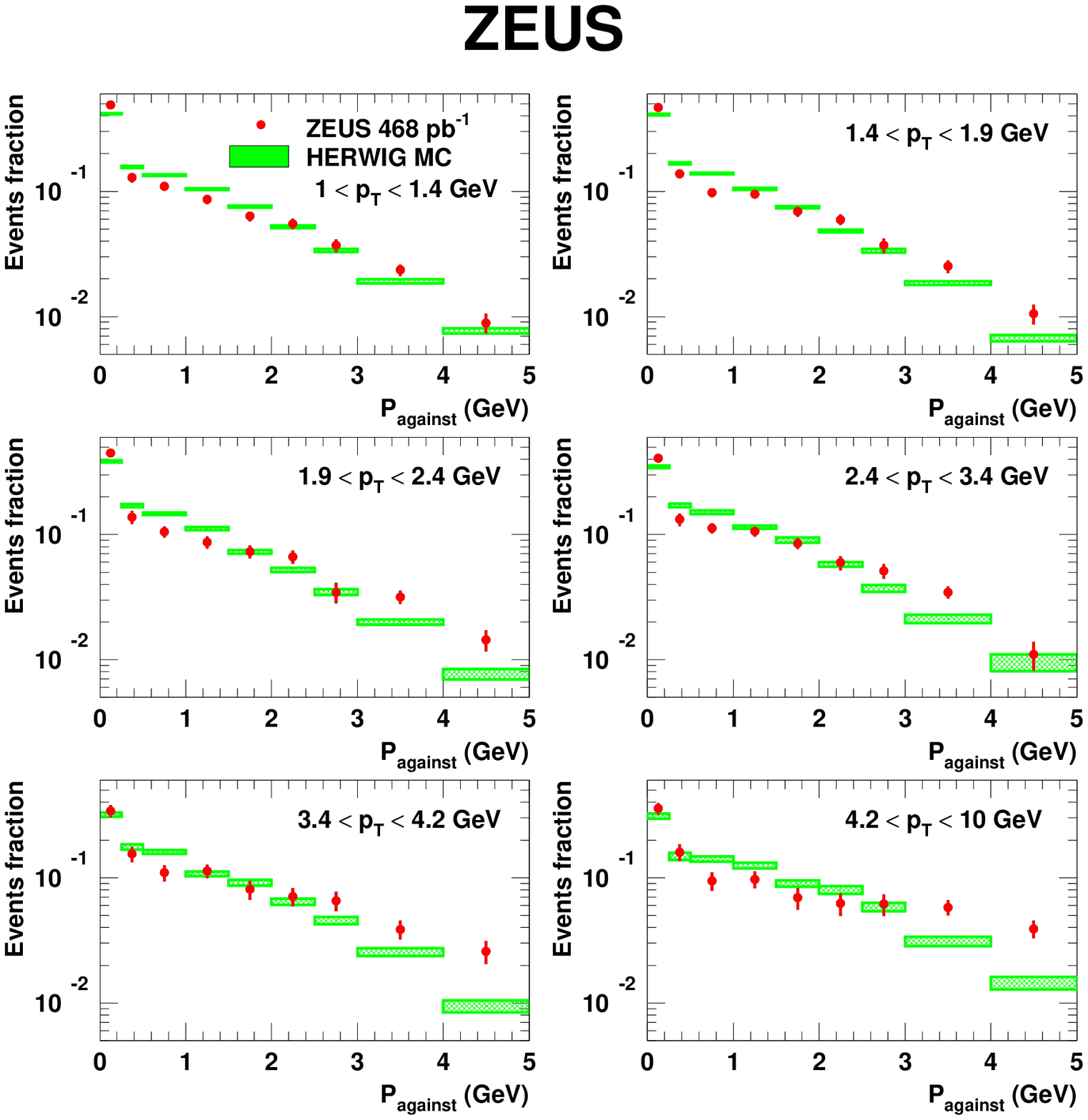}
\caption{Momentum flow against the \(J/\psi\) direction of flight
in the laboratory frame, \pagainst, for different $p_T$ ranges. 
The distributions are normalized to unity and are not corrected for
detector acceptance.
The measurement is performed in the kinematic region 
$60 < W < 240$ GeV and $0.3 < z < 0.9$. 
The data are shown as points with error bars indicating their uncertainties.
The predictions obtained from the {\sc Herwig} MC are also shown as 
rectangular shaded boxes. The height of these boxes represents the
uncertainties of the prediction.}
\label{fig:pagainst}
\end{figure}

\begin{figure}[hbpt!]
\unitlength1cm
\hspace{-0.75cm}
\includegraphics[width=1.05\textwidth]{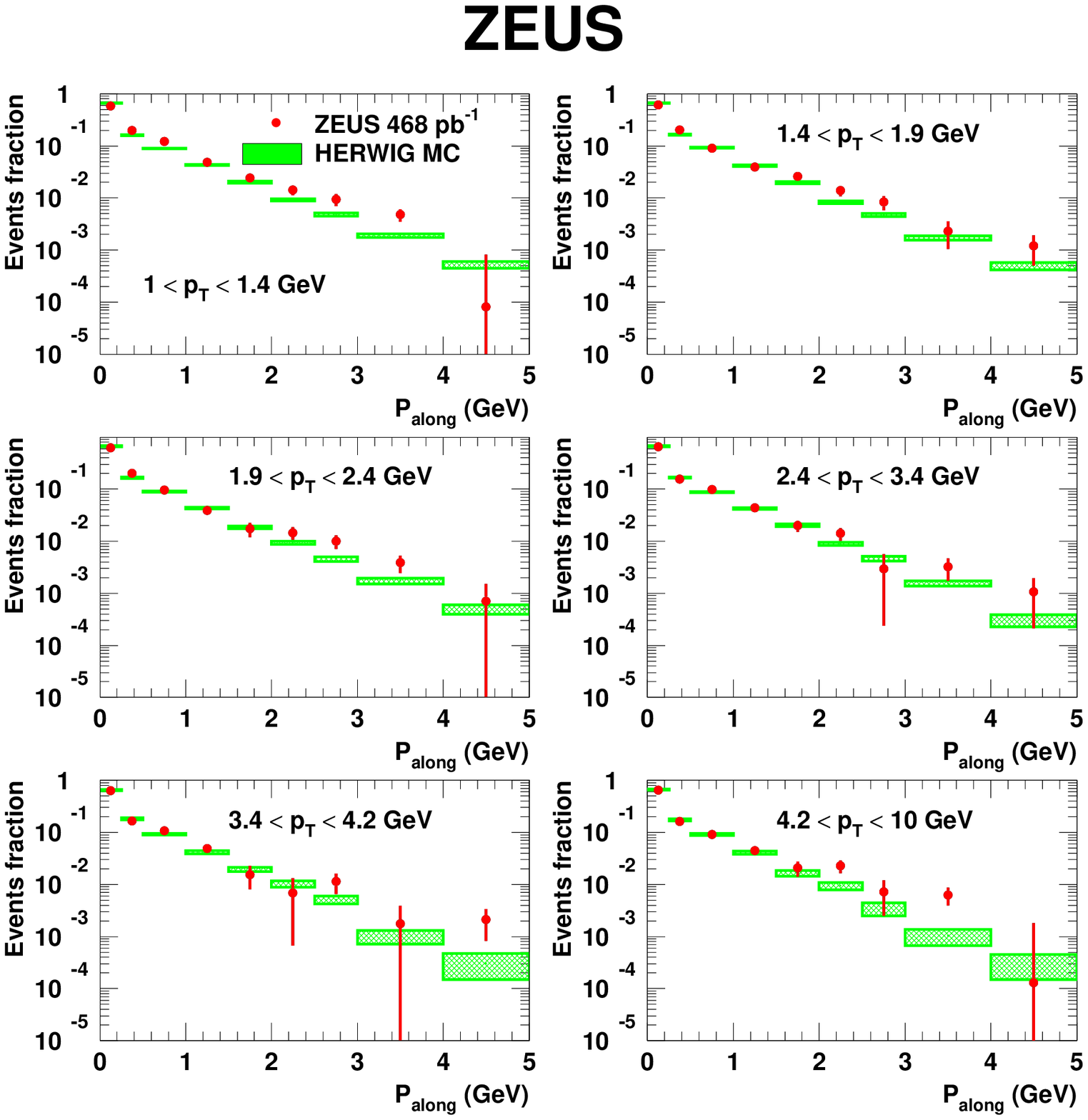}
\caption{Momentum flow along the \(J/\psi\) direction of flight
in the laboratory frame, \palong, for different $p_T$ ranges. 
The distributions are normalized to unity and are not corrected for
detector acceptance.
The measurement is performed in the kinematic region 
$60 < W < 240$ GeV and $0.3 < z < 0.9$. 
The data are shown as points with error bars indicating their uncertainties.
The predictions obtained from the {\sc Herwig} MC are also shown as 
rectangular shaded boxes. The height of these boxes represents the
uncertainties of the prediction.}
\label{fig:palong}
\end{figure}